\pdfoutput=1
\documentclass[cernpreprint,texlive=2011,txfonts,UKenglish]{latex/atlasdoc}

\usepackage[subfigure=true,biblatex=false]{latex/atlaspackage}
\usepackage[BSM,hion,journal,math,misc,other,xref,process=true]{latex/atlasphysics}
\usepackage{latex/atlascontribute}

\usepackage{multirow}
\usepackage{rotating}
\usepackage{xtab}
\usepackage{cite}
\usepackage{siunitx}

\usepackage{common/jetetmisssymbols}
\usepackage{common/boostedsymbols}
\usepackage{common/edittools}
\usepackage{common/susysymbols}
\usepackage{common/atlas_rounding}

\newcommand{\intluminoerr}       {17.4\invfb}

\graphicspath{{logos/}{figures/}}
\newcommand{\papertitle}{
A search for top squarks with $\mathbf{\it R}$-parity-violating decays to all-hadronic final states with the ATLAS detector in $\mathbf{\sqrt{s}=8}$~TeV proton--proton collisions
}

\AtlasAbstract{%
A search for the pair production of top squarks, each with $R$-parity-violating decays into two Standard Model quarks, is performed using \intluminoerr of \sqseight proton--proton collision data recorded by the ATLAS experiment at the LHC. Each top squark is assumed to decay to a $b$- and an $s$-quark, leading to four quarks in the final state. Background discrimination is achieved with the use of $b$-tagging and selections on the mass and substructure of large-radius jets, providing sensitivity to top squark masses as low as 100~GeV. No evidence of an excess beyond the Standard Model background prediction is observed and top squarks decaying to $\bbar\sbar$ are excluded for top squark masses in the range $100 \leq \mstop \leq 315\GeV$ at 95\% confidence level. 
}

\AtlasRefCode{SUSY-2015-05}
\AtlasJournal{JHEP}

\AtlasCoverEditor{D. W. Miller}{David.W.Miller@cern.ch}
\AtlasCoverEditor{J. Olsson}{joakim.olsson@cern.ch}

\AtlasCoverSupportingNote{Supporting documentation}{https://cds.cern.ch/record/1978952}

\AtlasCoverAnalysisTeam{David Miller, Joakim Olsson}
\AtlasContributor{David Miller}{Primary editor, supervisor}
\AtlasContributor{Joakim Olsson}{Analysis, limit setting, data processing, plots}

\AtlasCoverEgroupEditors{atlas-susy-2015-05-editors@cern.ch}
\AtlasCoverEgroupEdBoard{atlas-susy-2015-05-editorial-board@cern.ch}

\AtlasCoverEdBoardMember{Michael Flowerdew (Chair, Munich MPI)}{}
\AtlasCoverEdBoardMember{Walter Hopkins (Oregon)}{}
\AtlasCoverEdBoardMember{Dirk Zerwas (Orsay LAL)}{}

\PreprintIdNumber{CERN-PH-EP-2015-296}
\AtlasDate{10th June 2016} 
\AtlasJournalRef{JHEP06 (2016) 067} 
\arXivId{1601.07453}
\AtlasDOI{10.1007/JHEP06(2016)067} 

\author{The ATLAS Collaboration}

\hypersetup{pdftitle={rpvstop-paper},pdfauthor={The ATLAS Collaboration},pageanchor=false}
\AtlasTitle{\papertitle}
                             
\begin{document}

\maketitle

\section{Introduction}
\label{sec:introduction}
Supersymmetry (\SUSY) is an extension of the Standard Model (SM)~\cite{Golfand:1971iw,Volkov:1973ix,Wess:1974tw,Wess:1974jb,Ferrara:1974pu,Salam:1974ig,Martin:1997ns} that fundamentally relates fermions and bosons. It is an especially alluring theoretical possibility given its potential to solve the hierarchy problem~\cite{Sakai:1981gr,Dimopoulos:1981yj,Ibanez:1981yh,Dimopoulos:1981zb} and to provide a dark-matter candidate~\cite{Goldberg:1983nd,Ellis:1983ew}. 

This paper presents a search for the pair production of supersymmetric top squarks (stops)\footnote{The superpartners of the left- and right-handed top quarks, \stopL and \stopR, mix to form the two mass eigenstates \stopone and \stoptwo, where \stopone is the lighter one. This analysis focuses on \stopone, which is referred to hereafter as \stop.}, which then each decay to two SM quarks, using \intluminoerr of \sqseight proton--proton (\pp) collision data recorded by the ATLAS experiment at the Large Hadron Collider (LHC). This decay violates the $R$-parity conservation (RPC)~\cite{Farrar:1978xj} assumed by most searches for stops~\cite{Aad:2015pfx,Khachatryan:2015wza}. In RPC scenarios, \SUSY particles are required to be produced in pairs and decay to the lightest supersymmetric particle (LSP), which is stable. In $R$-parity-violating (RPV) models, decays to only SM particles are allowed, and generally relax the strong constraints now placed on standard RPC \SUSY scenarios by the LHC experiments. It is therefore crucial to expand the scope of the \SUSY search programme to include RPV models. Common signatures used for RPV searches include resonant lepton-pair production~\cite{Aad:2015pfa}, exotic decays of long-lived particles with displaced vertices~\cite{Aad:2013gva, SUSY-2013-22, Aad:2015rba, Aad:2015qfa}, high lepton multiplicities~\cite{SUSY-2013-13, Chatrchyan:2013xsw}, and high-jet-multiplicity final states~\cite{Aad:2015lea}. Scenarios which have stops of mass below 1 TeV are of particular interest as these address the hierarchy problem~\cite{Inoue:1982pi,Ellis:1983ed,Barbieri:1987fn,deCarlos:1993yy}.

\SUSY RPV decays to SM quarks and leptons are controlled by three Yukawa couplings in the generic supersymmetric superpotential~\cite{Dreiner:1998wm, Allanach:2003eb}. These couplings are represented by $\lamijk , \lampijk , \lamppijk$, where $i,j,k\in{1,2,3}$ are generation indices that are sometimes omitted in the discussion that follows. The first two ($\lam , \lamp$) are lepton-number-violating couplings, whereas the third ($\lampp$) violates baryon number. It is therefore generally necessary that either of the couplings to quarks, \lamp or \lampp, be vanishingly small to prevent spontaneous proton decay~\cite{Martin:1997ns}. It is common to consider non-zero values of each coupling separately.
Scenarios in which $\lampp \neq 0$ are often referred to UDD scenarios because of the baryon-number-violating term that \lampp controls in the superpotential. 
Current indirect experimental constraints~\cite{Allanach:1999ic} on the sizes of each of the UDD couplings \lampp from sources other than proton decay are primarily valid for low squark mass and for first- and second-generation couplings. Those limits are driven by double nucleon decay~\cite{Sher:1994sp} (for $\lampp_{112}$), neutron oscillations~\cite{Zwirner1983103} (for $\lampp_{113}$), and \Zboson-boson branching ratios~\cite{Bhattacharyya:1997vv}.

\begin{figure}[!ht]
  \centering
  \includegraphics[width=0.4\columnwidth]{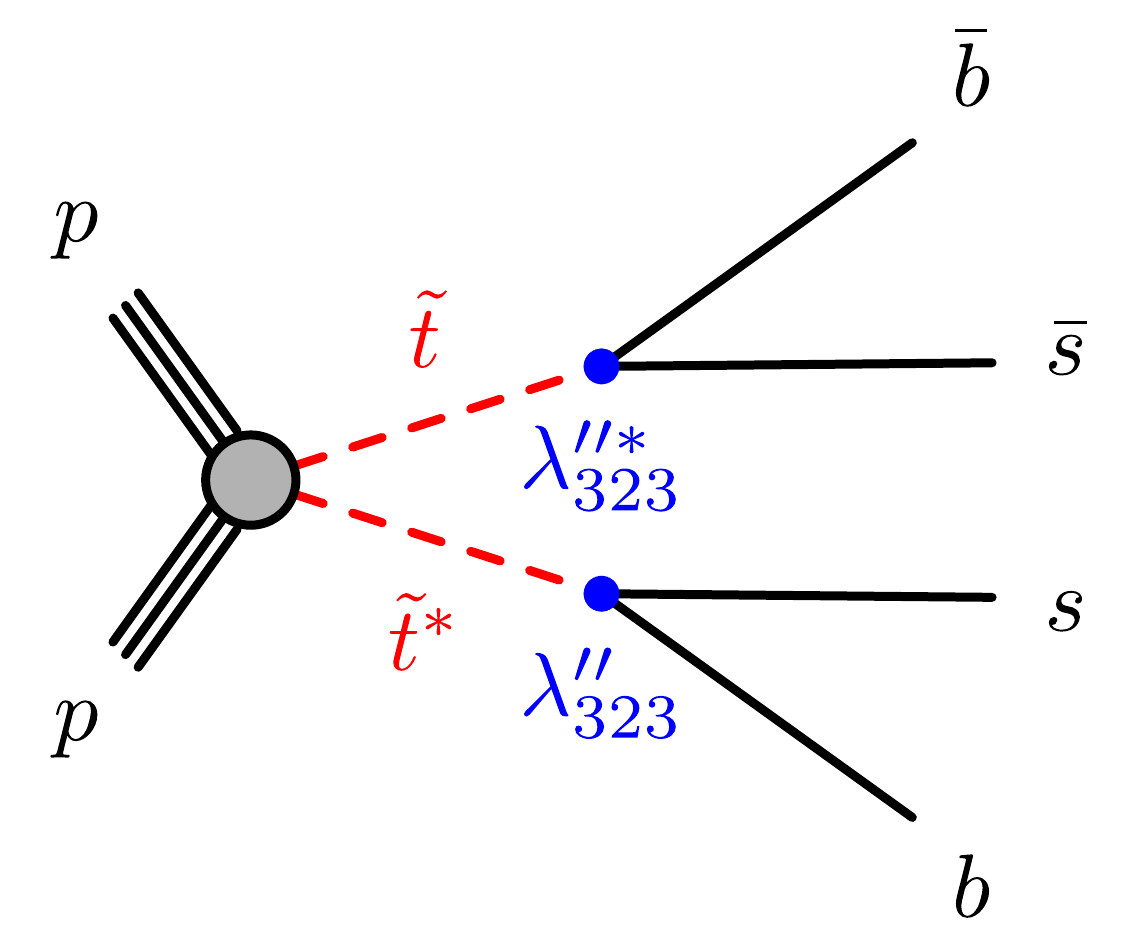}
  \caption{Benchmark signal process considered in this analysis. The  solid black lines represent Standard Model particles, the dashed red lines represent the stops, and the blue points represent RPV vertices labelled by the relevant coupling for this diagram.}
  \label{fig:intro:diagram}
\end{figure}

The benchmark model considered in this paper is a baryon-number-violating RPV scenario in which the stop is the LSP. The search specifically targets low-mass stops in the range 100--400$\GeV$ that decay via the $\lampp_{323}$ coupling, thus resulting in stop decays $\stop\ra\bbar\sbar$ (assuming a $100\%$ branching ratio) as shown in \Figref{intro:diagram}. The motivation to focus on the third-generation UDD coupling originates primarily from the minimal flavour violation (MFV) hypothesis~\cite{DAmbrosio:2002ex} and the potential for this decay channel to yield a possible signal of RPV \SUSY with a viable dark-matter candidate~\cite{Batell:2013zwa}. The MFV hypothesis essentially requires that all flavour- and CP-violating interactions are linked to the known structure of Yukawa couplings, and has been used to argue for the importance of the \lampp couplings~\cite{Csaki:2011ge}.  

The process $\stop \stop^{\ast} \rightarrow \bbar \sbar b s$ represents an important channel in which to search for SUSY in scenarios not yet excluded by LHC data~\cite{Csaki:2011ge, Batell:2013zwa, Bai:2013xla}. Some of the best constraints on this process are from the ALEPH Collaboration, which set lower bounds on the mass of the stop at $\mstop \gtrsim 80$~GeV~\cite{LEPRPVStop}. The CDF Collaboration extended these limits, excluding $50 \lesssim \mstop \lesssim 90$~GeV~\cite{Aaltonen:2013hya}. The CMS Collaboration recently released the results of a search that excludes $200 \lesssim \mstop \lesssim 385$~GeV~\cite{Khachatryan:2014lpa} in the case where heavy-flavour jets are present in the final state. In addition, two ATLAS searches have placed constraints on RPV stops that decay to $\bar{b} \bar{s}$ when they are produced in the decays of light gluinos ($\mgluino \lesssim\,$900--1000$\GeV$)~\cite{SUSY-2013-04, SUSY-2013-09}. The search presented here specifically focuses on direct stop pair production and seeks to close the gap in excluded stop mass between $\sim100$--$200\GeV$. 
Contributions from RPV interactions at production -- such as would be required for resonant single stop production -- are neglected in this analysis. This approach is valid provided that the RPV interaction strength is small compared to the strong coupling constant, which is the case for $\lampp_{323} \lesssim 10^{-2}$--$10^{-1}$~\cite{Barbier:2004ez} and for the estimated size of $\lampp_{323} \sim 10^{-4}$ from MFV in the model described in Ref.~\cite{Csaki:2011ge}.

The reduced sensitivity of standard \SUSY searches to RPV scenarios is primarily due to the limited effectiveness of the high missing transverse momentum requirements used in the event selection common to many of those searches, motivated by the assumed presence of undetected LSPs. Consequently, the primary challenge in searches for RPV \SUSY final states is to identify suitable substitutes for background rejection to the canonical large missing transverse momentum signature.   

Backgrounds dominated by multijet final states typically overwhelm the signal in the four-jet topology.
In order to overcome this challenge, new observables are employed to search for $\stop \stop^{\ast} \rightarrow \bbar \sbar b s$ in the low-\mstop regime~\cite{Bai:2013xla}. For $\mstop \approx100$--$300\GeV$, the initial stop transverse momentum (\pt) spectrum extends significantly into the range for which $\pT \gg \mstop$. This feature is the result of boosts received from initial-state radiation (ISR) as well as originating from the parton distribution functions (PDFs). As the Lorentz boost of each stop becomes large, the stop decay products begin to merge with a radius roughly given by $\DeltaR \approx 2\mstop/\pT$, and thus can be clustered together within a single large-radius (\largeR) jet with a mass $\massjet \approx \mstop$. By focusing on such cases, the dijet and multijet background can be significantly reduced via selections that exploit this kinematic relationship and the structure of the resulting \textit{stop jet}, in a similar way to boosted objects used in previous measurements and searches by ATLAS~\cite{STDM-2011-19, STDM-2011-38, TOPQ-2011-23, PERF-2012-02, STDM-2012-21}. In this case, since the stop is directly produced in pairs instead of from the decay of a massive parent particle, the strategy is most effective at low \mstop where the boosts are the largest.

\section{The ATLAS detector}
\label{sec:detector}
The ATLAS detector~\cite{detPaper,PerfWithData2010} provides nearly full solid angle\footnote{The ATLAS reference system is a Cartesian right-handed coordinate system, with the nominal collision point at the origin. The anticlockwise beam direction defines the positive $z$-axis, while the positive $x$-axis is defined as pointing from the collision point to the centre of the LHC ring and the positive $y$-axis points upwards. The azimuthal angle $\phi$ is measured around the beam axis, and the polar angle $\theta$ is measured with respect to the $z$-axis. Pseudorapidity is defined as $\eta = - \ln [\tan(\theta/2)]$, rapidity is defined as $y = 0.5\ \ln[(E + p_z)/(E - p_z)]$, where $E$ is the energy and $p_z$ is the $z$-component of the momentum, and transverse energy is defined as $\et = E \sin \theta$.} coverage around the collision point with an inner tracking system (inner detector, or ID) covering the pseudorapidity range $|\eta|<2.5$,  electromagnetic (EM) and hadronic calorimeters covering $|\eta|<4.9$, and a muon spectrometer covering $|\eta|<2.7$ that provides muon trigger capability up to $|\eta|<2.4$.

The ID comprises a silicon pixel tracker closest to the beamline, a microstrip silicon tracker, and a straw-tube transition-radiation tracker at radii up to $108$ cm. A thin solenoid surrounding the tracker provides a $2~\rm T$ axial magnetic field enabling the measurement of charged-particle momenta. The overall ID acceptance spans the full azimuthal range in $\phi$, and the range $|\eta|<2.5$ for particles originating near the nominal LHC interaction region~\cite{STDM-2010-06}.

The EM and hadronic calorimeters are composed of multiple subdetectors spanning $|\eta|\leq4.9$. The EM barrel calorimeter uses a  liquid-argon (LAr) active medium and lead absorbers. In the region $|\eta| < 1.7$, the hadronic (Tile) calorimeter is constructed from steel absorber and scintillator tiles and is separated into barrel ($|\eta|<1.0$) and extended-barrel ($0.8<|\eta|<1.7$) sections. The endcap ($1.375<|\eta|<3.2$) and forward ($3.1<|\eta|<4.9$) regions are instrumented with LAr calorimeters for EM as well as hadronic energy measurements.

A three-level trigger system is used to select events to record for offline analysis. The different parts of the trigger system are referred to as the level-1 trigger, the level-2 trigger, and the event filter~\cite{PERF-2011-02}. The level-1 trigger is implemented in hardware and uses a subset of detector information to reduce the event rate to a design value of at most 75~kHz. The level-1 trigger is followed by two software-based triggers, the level-2 trigger and the event filter, which together reduce the event rate to a few hundred Hz. The search presented in this document uses a trigger that requires a high-\pt \ jet and a large summed jet transverse momentum (\HT), as described in \secref{event-selection}.

\section{Monte Carlo simulation samples}
\label{sec:mc}
Monte Carlo (MC) simulation is used to study the signal acceptance and systematic uncertainties, to test the background estimation methods used, and to estimate the \ttbar background. In all cases, events are passed through the full \geant~\cite{Geant4} detector simulation of ATLAS~\cite{simulation} after the simulation of the parton shower and hadronisation processes. Following the detector simulation, identical event reconstruction and selection criteria are applied to both the MC simulation and to the data. Multiple \pp collisions in the same and neighbouring bunch crossings (\pileup) are simulated for all samples by overlaying additional soft \pp collisions which are  generated with \Pythia 8.160~\cite{pythia8} using the $\mbox{ATLAS A2}$ set of tuned parameters (tune) in the MC generator~\cite{ATL-PHYS-PUB-2011-014} and the MSTW2008LO PDF set~\cite{PDF-MSTW2008}. These additional interactions are overlaid onto the hard scatter and events are reweighted such that the MC distribution of the average number of \pp interactions per bunch crossing matches the measured distribution in the full 8~TeV data sample. 

The signal process is simulated using \Herwigpp 2.6.3a~\cite{Herwigpp} with the UEEE3 tune~\cite{Gieseke:2012ft} for several stop-mass hypotheses using the PDF set CTEQ6L1~\cite{PDF-CTEQ, cteq6l1}. All non-SM particles masses are set to 5~TeV except for the stop mass, which is scanned in $25\GeV$ steps from $\mstop = 100\GeV$ to $\mstop = 400\GeV$. 

\begin{figure}[!ht]
  \centering
  \includegraphics[width=0.8\columnwidth]{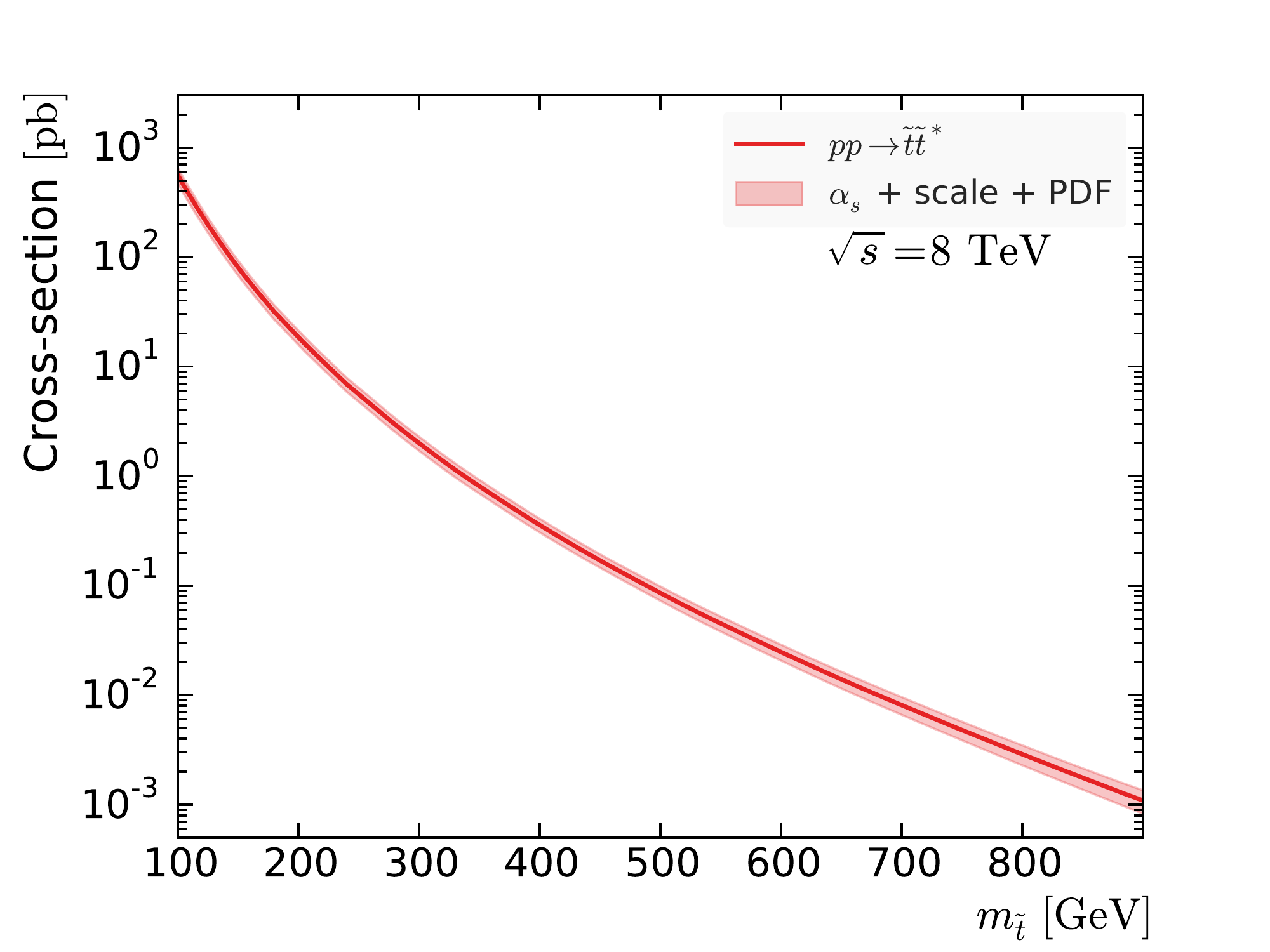}
  \caption{Cross-section for direct $\stop\stop^{*}$ pair production at the LHC centre-of-mass energy of \sqseight~\cite{Beenakker:1997ut,Beenakker:2010nq,Beenakker:2011fu}.}
  \label{fig:intro:xsect}
\end{figure}

The signal cross-section used (shown in \figref{intro:xsect}) is calculated to next-to-leading order in the strong coupling constant, adding the resummation of soft gluon emission at next-to-leading-logarithmic accuracy (NLO+NLL)~\cite{Beenakker:1997ut,Beenakker:2010nq,Beenakker:2011fu}. For the range of stop masses considered, the uncertainty on the cross-section is approximately 15\%~\cite{Kramer:2012bx}. \Madgraph 5.1.4.8~\cite{Alwall:2011uj} is used to study the impact of ISR on the stop \pt spectrum. The \Madgraph samples have  one additional parton in the matrix element, which improves the modelling of a hard ISR jet. \Madgraph is then interfaced to \Pythia 6.426 with the AUET2B tune~\cite{ATL-PHYS-PUB-2011-009} and the CTEQ6L1 PDF set for parton shower and hadronisation. The distribution of $\pt(\stop\stop^*)$ from the nominal \Herwigpp signal sample is then reweighted to match that of the \texttt{MadGraph+PYTHIA} sample.

Dijet and multijet events, as well as top quark pair (\ttbar) production processes, are simulated in order to study the SM contributions and background estimation techniques. For optimisation studies, SM dijet and multijet events are generated using \Herwigpp 2.6.3a with the CTEQ6L1 PDF set. Top quark pair events are generated with the \Powheg-BOX-r2129 \ \cite{Nason:2004rx, Frixione:2007vw, Alioli:2010xd} event generator with the CT10 NLO PDF set~\cite{Lai:2010vv}. These events are then interfaced to  \Pythia 6.426 with the P2011C tune~\cite{Perugia2010} and the same CTEQ6L1 PDF set as \Herwigpp. 

The \ttbar \ production cross-section is calculated at next-to-next-to-leading order (NNLO) in QCD including resummation of next-to-next-to-leading logarithmic (NNLL) soft gluon terms with \texttt{top++2.0} \cite{Czakon:2013goa, Czakon:2012pz, Czakon:2012zr, Baernreuther:2012ws, Cacciari:2011hy, Czakon:2011xx}. The value of the \ttbar \ cross-section is $\sigma_{\ttbar}= 253^{+13}_{-15}{\rm~pb}$.

\newpage

\section{Object definitions}
\label{sec:objects}
The data are required to have satisfied criteria designed to reject events with significant contamination from detector noise, non-collision beam backgrounds, cosmic rays, and other spurious effects. 
To reject non-collision beam backgrounds and cosmic rays, events are required to contain a primary vertex consistent with the LHC beamspot, reconstructed from at least two tracks with transverse momenta $\pttrk>400$~MeV. 
If more than one vertex satisfies these criteria, the primary vertex is chosen as the one with the highest $\sum_{\rm tracks}(\pt^2)$. 

The \akt algorithm~\cite{Cacciari:2008gp}, with a radius parameter of $R=0.4$, is used for initial jet-finding using version 3 of \mbox{\Fastjet}~\cite{Cacciari:2011ma}. 
The inputs to the jet reconstruction are three-dimensional \topos~\cite{TopoClusters}. This method first clusters together topologically connected calorimeter cells and classifies these clusters as either electromagnetic or hadronic. The classification uses a local cluster weighting calibration scheme based on cell-energy density and shower depth within the calorimeter~\cite{PERF-2012-01}.
Based on this classification, energy corrections are applied which are derived from single-pion MC simulations. Dedicated hadronic corrections are derived to account for the effects of differences in response to hadrons compared to electrons, signal losses due to noise-suppression threshold effects, and energy lost in non-instrumented regions. 
The final jet energy calibration is derived from MC simulation as a correction relating the calorimeter response to the jet energy at generator level. In order to determine these corrections, the same jet definition used in the reconstruction is applied to stable (with lifetimes greater than 10~ps) generator-level particles, excluding muons and neutrinos. A subtraction procedure is also applied in order to mitigate the effects of \pileup~\cite{Cacciari:2007fd}. Finally, the $R=0.4$ jets are further calibrated with additional correction factors derived \insitu from a combination of $\gamma$+jet, $Z$+jet, and dijet-balance methods~\cite{PERF-2012-01}.

All jets reconstructed with the \akt algorithm using a radius parameter of $R=0.4$ and a measured $\ptjet>20$~GeV are required to satisfy the quality criteria discussed in detail in Ref.~\cite{DAPR-2012-01}. 
These quality criteria selections for jets are extended to prevent contamination from detector noise through several detector-region-specific requirements. Jets contaminated by energy deposits due to noise in the forward hadronic endcap calorimeter are rejected and jets in the central region ($|\eta|<2.0$) that are at least 95\% contained within the EM calorimeter are required to not exhibit any electronic pulse shape anomalies~\cite{LARG-2013-01}. Any event with a jet that fails these requirements is removed from the analysis.

Identification of jets containing $b$-hadrons (so-called $b$-jets) is achieved through the use of a multivariate $b$-tagging algorithm referred to as MV1~\cite{Aad:2015ydr}. This algorithm is based on an artificial neural-network algorithm that exploits the impact parameters of charged-particle tracks, the parameters of reconstructed secondary vertices, and the topology of $b$- and $c$-hadron decays inside an \akt $R=0.4$ jet. A working point corresponding to a $70\%$ $b$-jet efficiency in simulated \ttbar events is used. The corresponding mis-tag rates, defined as the fraction of jets originating from non-$b$-jets which are tagged by the $b$-tagging algorithm in an inclusive jet sample, for light jets and $c$-jets are approximately $1\%$ and $20\%$, respectively.  To account for differences with respect to data, data-derived corrections are applied to the MC simulation for the identification efficiency of $b$-jets and the probability to mis-identify jets resulting from light-flavour quarks, charm quarks, and gluons.

Initial jet-finding is extended using an approach called \textit{jet re-clustering}~\cite{Nachman:2014kla}. This allows the use of larger-radius jet algorithms while maintaining the calibrations and systematic uncertainties associated with the input jet definition. Small-radius \akt $R=0.4$ jets with $\pt>20\GeV$ and $|\eta|<2.4$ are used as input without modification to an \akt $R=1.5$ \largeR jet algorithm, to identify the hadronic stop decays. The small-$R$ jets with $\pt < 50\GeV$ are required to have a jet vertex fraction (\JVF) of at least 50\%. 
After summing the $\pt$ of charged-particle tracks matched to a jet, the JVF is the fraction due to tracks from the selected hard-scattering interaction and it provides a means by which to suppress jets from \pileup.  

To further improve the background rejection, a splitting procedure is performed on each of the two leading \largeR jets. After jet-finding, the constituents of these \largeR jets -- the  \akt $R=0.4$ input objects -- are processed separately by the \ca (\CamKt) algorithm~\cite{Dokshitzer:1997in, Wobisch:1998wt}, as implemented in \mbox{\Fastjet3}. The \CamKt algorithm performs pair-wise recombinations of proto-jets (the inputs to the jet algorithm) purely based on their angular separation. Smaller-angle pairs are recombined first, thus the final recombined pair typically has the largest separation. The \CamKt final clustering is then undone by one step, such that there are two branches ''{\it a}'' and ''{\it b}''. 
The following \textit{splitting} criteria are then applied to the branches ''{\it a}'' and ''{\it b}'' of each of the two leading \largeR \ jets:
\begin{itemize}
  \item Both branches carry appreciable \pT\ relative to the \largeR jet: 
  \begin{equation}
    \label{eq:splitting:asym}
    \frac{\mathrm{min}[\pT(a),\pT(b)]}{\pT(\mathrm{large-}R)} > 0.1.
  \end{equation}
  \item The mass of each branch is small relative to its \pt: 
  \begin{equation}
    \label{eq:splitting:mpt}
    \frac{m(a)}{\pT(a)} < 0.3 \qquad \mathrm{and} \qquad \frac{m(b)}{\pT(b)} < 0.3.
  \end{equation}
\end{itemize}
%
If either of the leading two \largeR jets fails these selections, the event is discarded. This implementation is identical to Ref.~\cite{Bai:2013xla}, which is derived from the diboson-jet tagger~\cite{Son:2012mb}. This approach differs somewhat from that used in Ref.~\cite{Butterworth:2008iy} in that no requirement is placed on the relative masses of the \largeR and small-$R$ jets.

\section{Trigger and offline event selections}
\label{sec:event-selection}
Events must satisfy jet and \HT selections applied in the trigger which require $\HT = \sum \pt > 500\GeV$, calculated as the sum of level-2 trigger jets within $|\eta| < 3.2$, and a leading jet within $|\eta|<3.2$ with $\pt > 145\GeV$. This relatively low-threshold jet trigger came online part-way through the data-taking period in 2012 and collected \intluminoerr of data.  
The corresponding offline selections require events to have at least one \akt $R=0.4$ jet with $\pt > 175$~GeV and $|\eta|<2.4$, as well as $\HT > 650\GeV$, where the sum is over all \akt $R=0.4$ jets with $\pt>20\GeV$, $|\eta|<2.4$, and $\JVF>0.5$ if $\pt<50\GeV$. The cumulative trigger selection efficiency is greater than 99\% for these offline requirements. The offline event preselection further requires that at least two \largeR jets with $\pt>200\GeV$ and mass $>20\GeV$ be present in each event. These requirements select a range of phase space for low stop masses in which the transverse momentum of the stops is often significantly greater than their mass. 

The signal region (SR) is defined to suppress the large multijet background and to enhance the fraction of events that contain \largeR jets consistent with the production of stop pairs, with each stop decaying to a light quark and a $b$-quark. Simulation studies indicate that three kinematic observables are particularly useful for background discrimination: 

\begin{enumerate}
  \item The mass asymmetry between the two leading \largeR jets in the event (with masses $m_1$ and $m_2$, respectively), defined as
\begin{equation}
   \Asym=\frac{|m_1-m_2|}{m_1+m_2},
\end{equation}
differentiates signal from background since the two stop subjet-pair resonances are expected to be of equal mass. 

\item The (absolute value of the cosine of the) stop-pair production angle, $\costhstar$, with respect to the beam line in the centre-of-mass reference frame\footnote{This scattering angle, $\theta^*$, is formed by boosting the two stop \largeR jets to the centre-of-mass frame and measuring the angle of either stop \largeR jet with respect to the beam line.} distinguishes between centrally produced massive particles and high-mass forward-scattering events from QCD. It provides efficient discrimination and does not exhibit significant variation with the stop mass. 

\item In addition, a requirement on the subjets is applied to each of the leading \largeR jets in the event.  The \pt\ of each subjet \textit{a} and \textit{b} relative to the other is referred to as the \ptratio, defined by 
\begin{equation}
   \ptratio = \frac{\mathrm{min}[\pt(a),\pt(b)]}{\mathrm{max}[\pt(a),\pt(b)]}.
\end{equation}
\end{enumerate}
The $\Asym$, $\costhstar$, and $\ptratio$ variables provide good discrimination between signal and background and are motivated by an ATLAS search for scalar gluons at $\sqrt{s}=7$~TeV~\cite{SUSYsgluon2010} as well as by Refs.~\cite{Bai:2013xla, Schumann:2011ji}.

\begin{figure}[!ht]
    \centering
    \subfigure[]{
      \includegraphics[width=0.48\columnwidth]{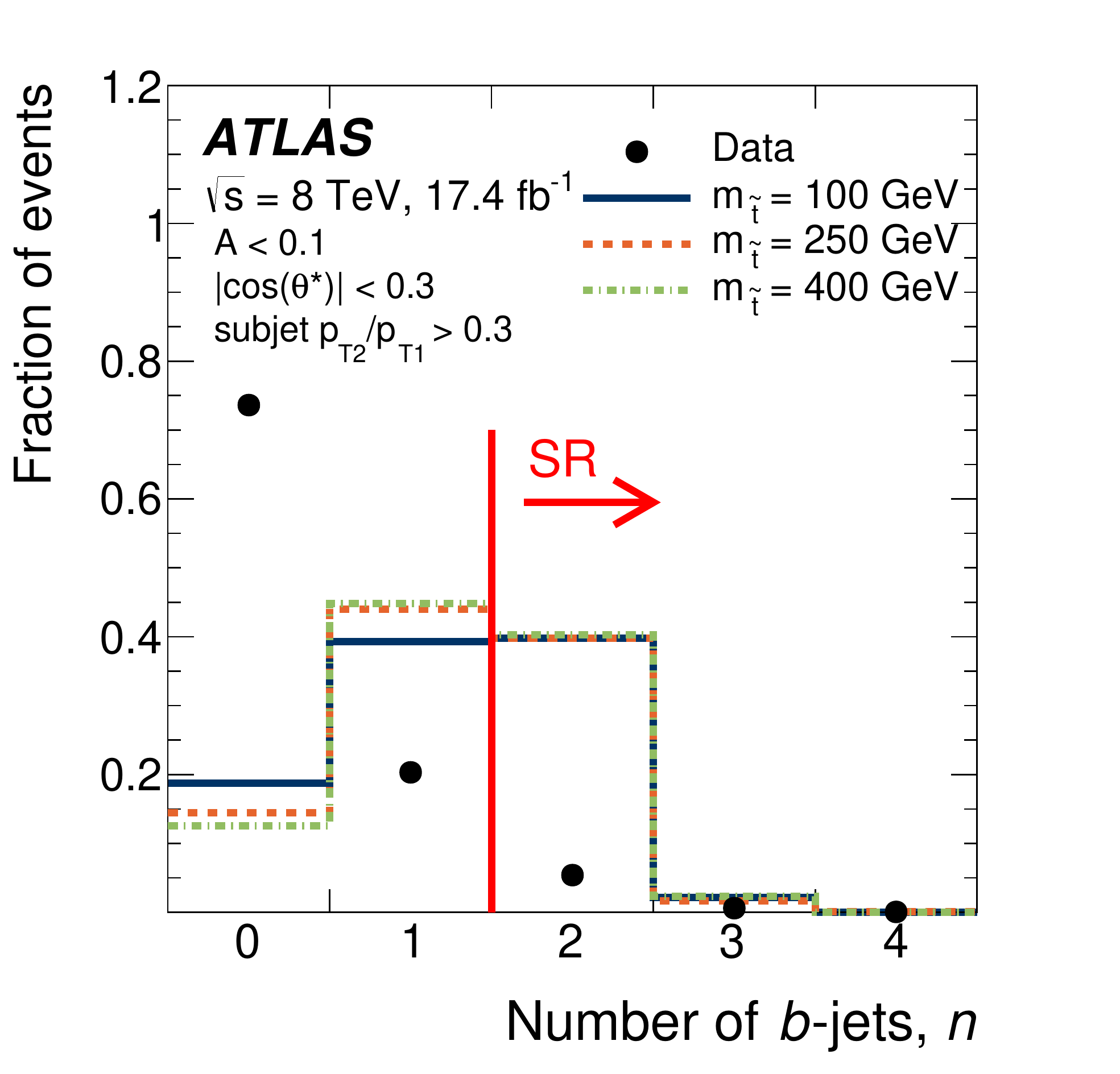}
      \label{fig:sigbkg:nbtags}
    }
    \subfigure[]{
     \includegraphics[width=0.48\columnwidth]{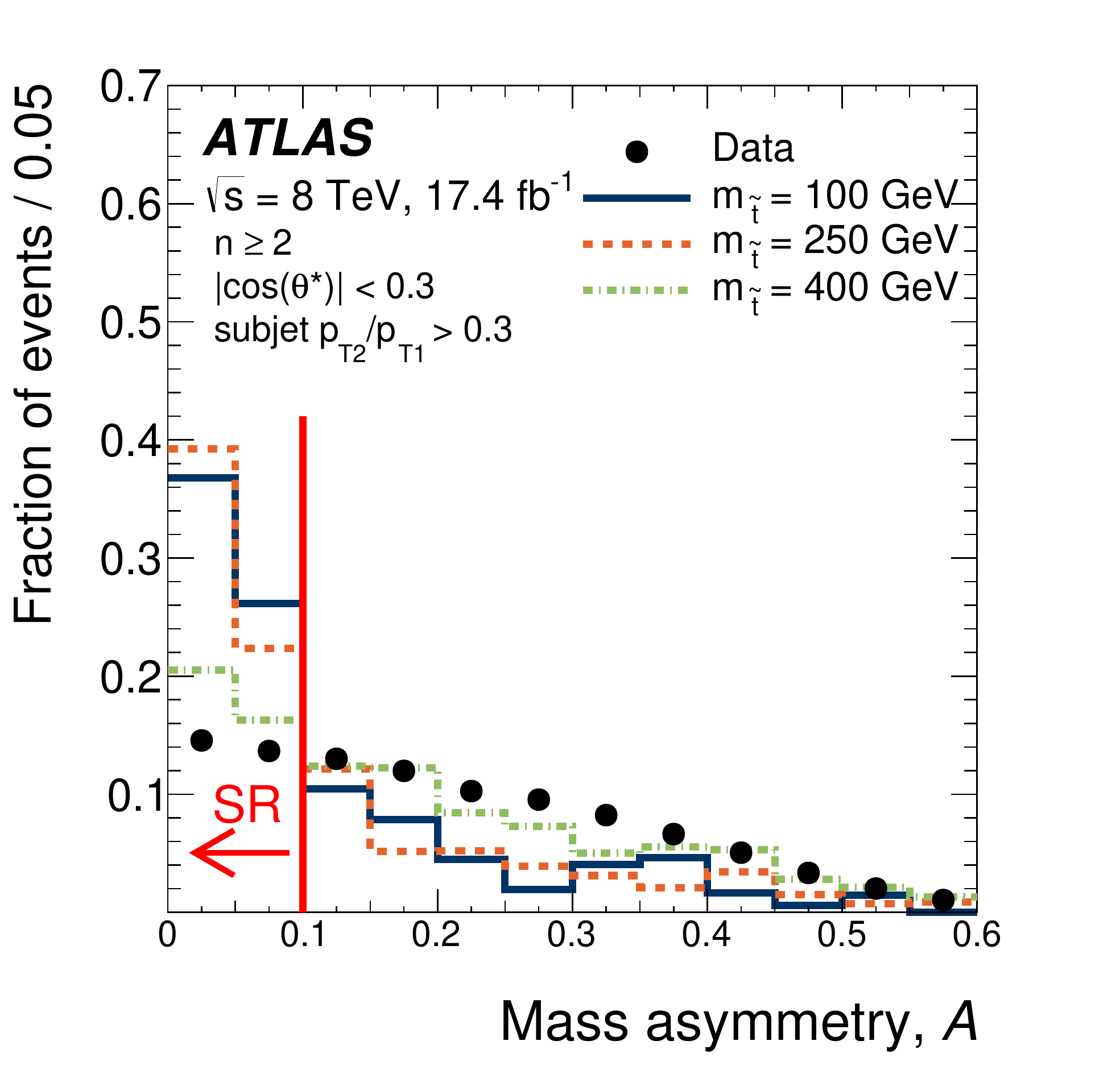}
      \label{fig:sigbkg:asym}
    } \\
    \subfigure[]{
     \includegraphics[width=0.48\columnwidth]{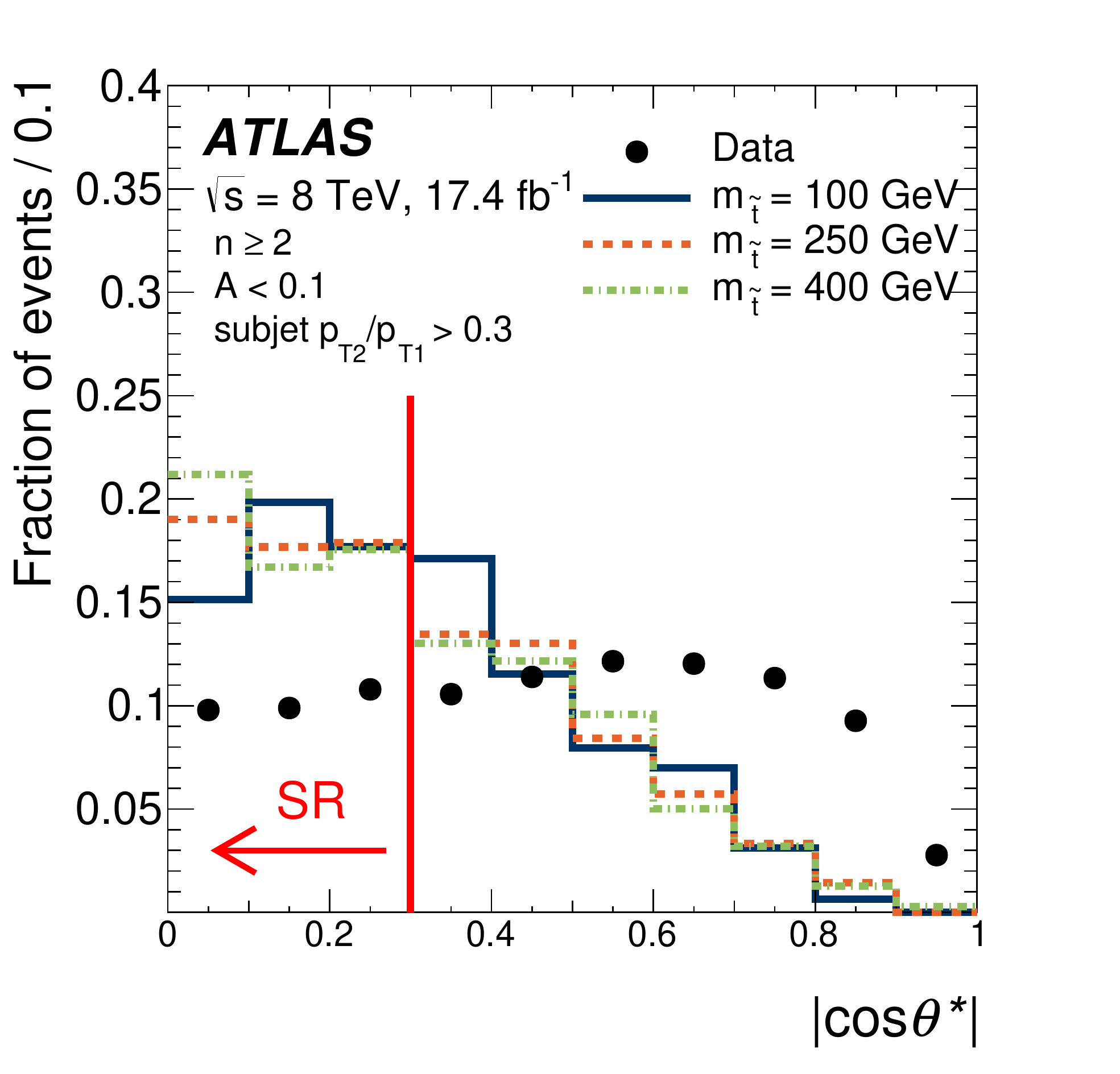}
      \label{fig:sigbkg:costhstar}
    }
    \subfigure[]{
     \includegraphics[width=0.48\columnwidth]{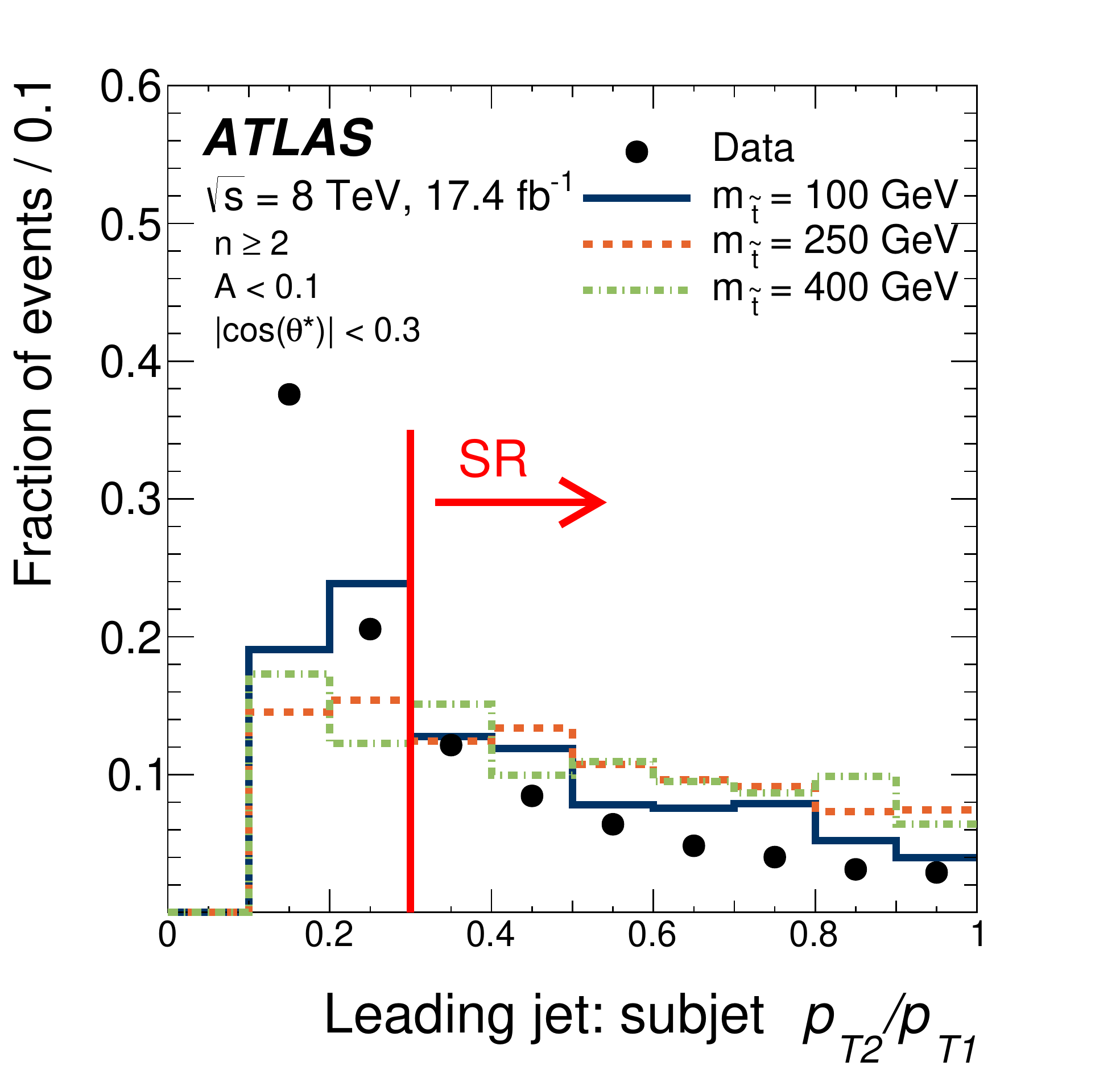}
      \label{fig:sigbkg:subjetpt}
    }
    \caption{Distributions of the discriminating variables for events in which the other three selections are applied for each subfigure. The signal region is indicated with a red arrow. All distributions are normalised to unity. Overflows are included in the last bin for subfigures \subref{fig:sigbkg:nbtags} and \subref{fig:sigbkg:asym}. \subref{fig:sigbkg:nbtags} Number of $b$-tags/event, $\nbtag$. \subref{fig:sigbkg:asym} Large-$R$ jet mass asymmetry, $\Asym$. \subref{fig:sigbkg:costhstar} Stop-pair centre-of-mass frame production angle, $\costhstar$. \subref{fig:sigbkg:subjetpt} Subjet $p_{\rm T2}/p_{\rm T1}$ for the leading jet in each event.}
    \label{fig:sigbkg:discrimination}
\end{figure}

In addition to the kinematic observables described above, $b$-tagging applied to \akt $R=0.4$ jets provides a very powerful discriminant for defining both the signal and the control regions, and one that is approximately uncorrelated with the kinematic features discussed above. 
Using these kinematic observables and the presence of at least two $b$-tagged jets per event, the signal region is defined by (for the leading two \largeR jets)
\begin{align}
  \Asym &< 0.1, \nonumber \\
  \costhstar &< 0.3, \label{eq:SRcuts} \\
  \ptratio &> 0.3. \nonumber
\end{align}
Distributions of the discriminating variables are shown in \figref{sigbkg:discrimination}. Insofar as the data points are dominated by background in these plots, even in the case of a potential signal, the data points should be understood to represent the background. 

\begin{figure}[!ht]
    \centering
    \subfigure[Linear scale]{
     \includegraphics[width=0.72\columnwidth]{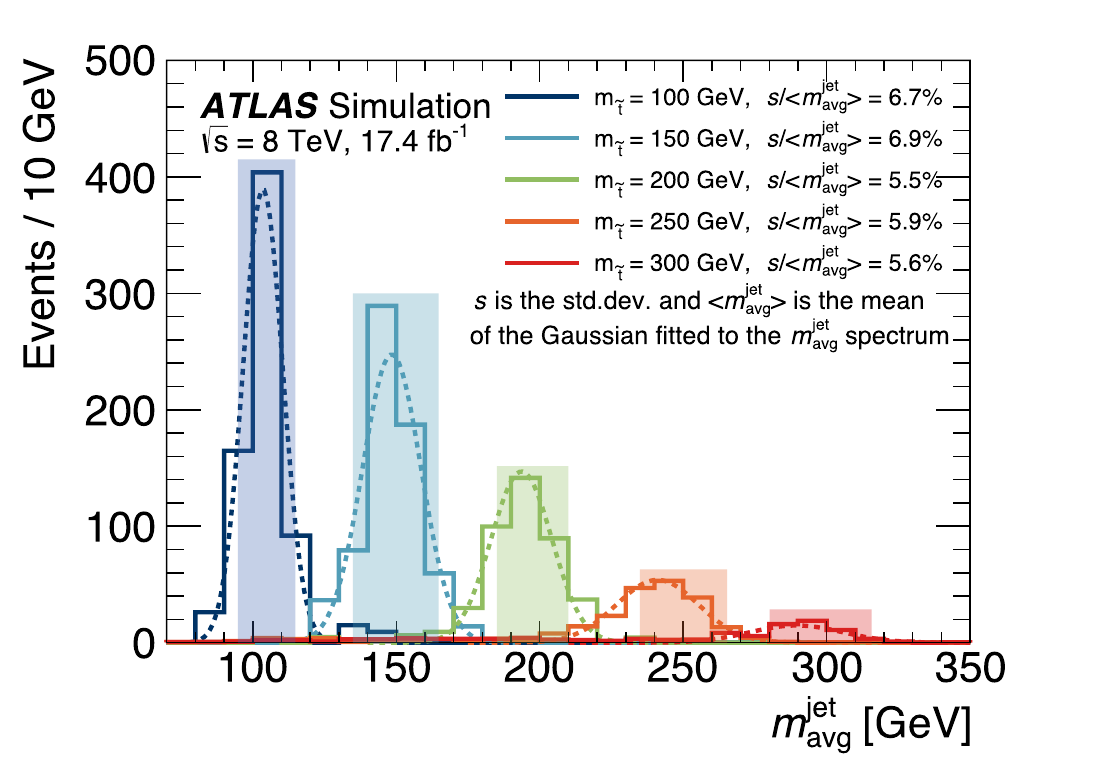}
      \label{fig:massres:linear}
    }
    \subfigure[Logarithmic scale]{
     \includegraphics[width=0.72\columnwidth]{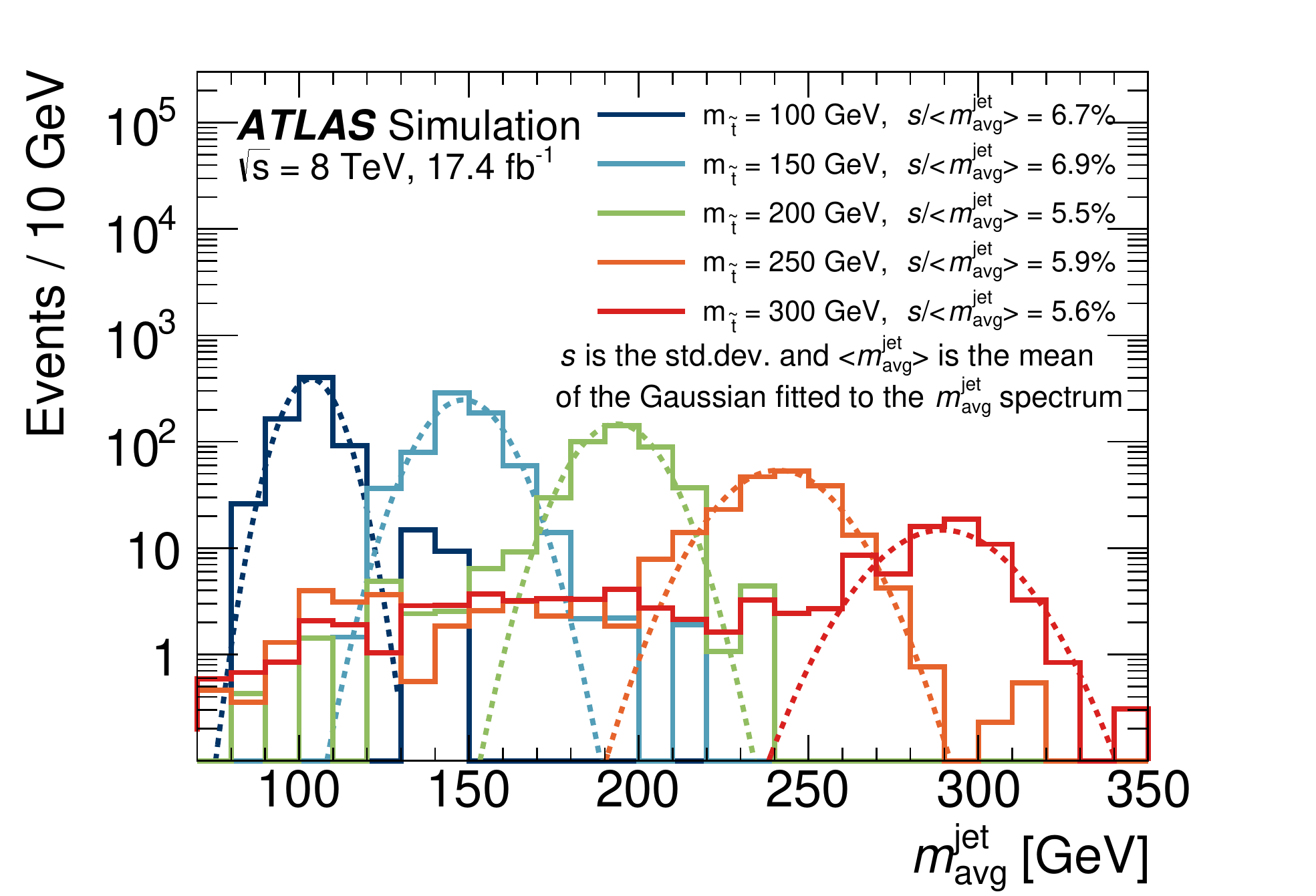}
      \label{fig:massres:logy}
    }
    \caption{Distributions of the average jet mass $\mavg$ for signal samples with $\mstop$ = 100, 150, 200, 250, and 300$\gev$, in linear \subref{fig:massres:linear} and logarithmic \subref{fig:massres:logy} scales (solid lines). A Gaussian distribution is fitted to the mass peak of each sample (dashed lines). The resolution, $s/\langle\mavg\rangle$, is quoted for each stop mass value. The mass windows are highlighted with the shaded rectangles in \subref{fig:massres:linear}. The long tail peaking around $\mstop/2$ for high-mass stops shown in \subref{fig:massres:logy} is due to events where not all stop decay products are clustered within the \largeR jets.}
    \label{fig:massres}
\end{figure}

\begin{figure}[!ht]
    \centering
     \includegraphics[width=0.6\columnwidth]{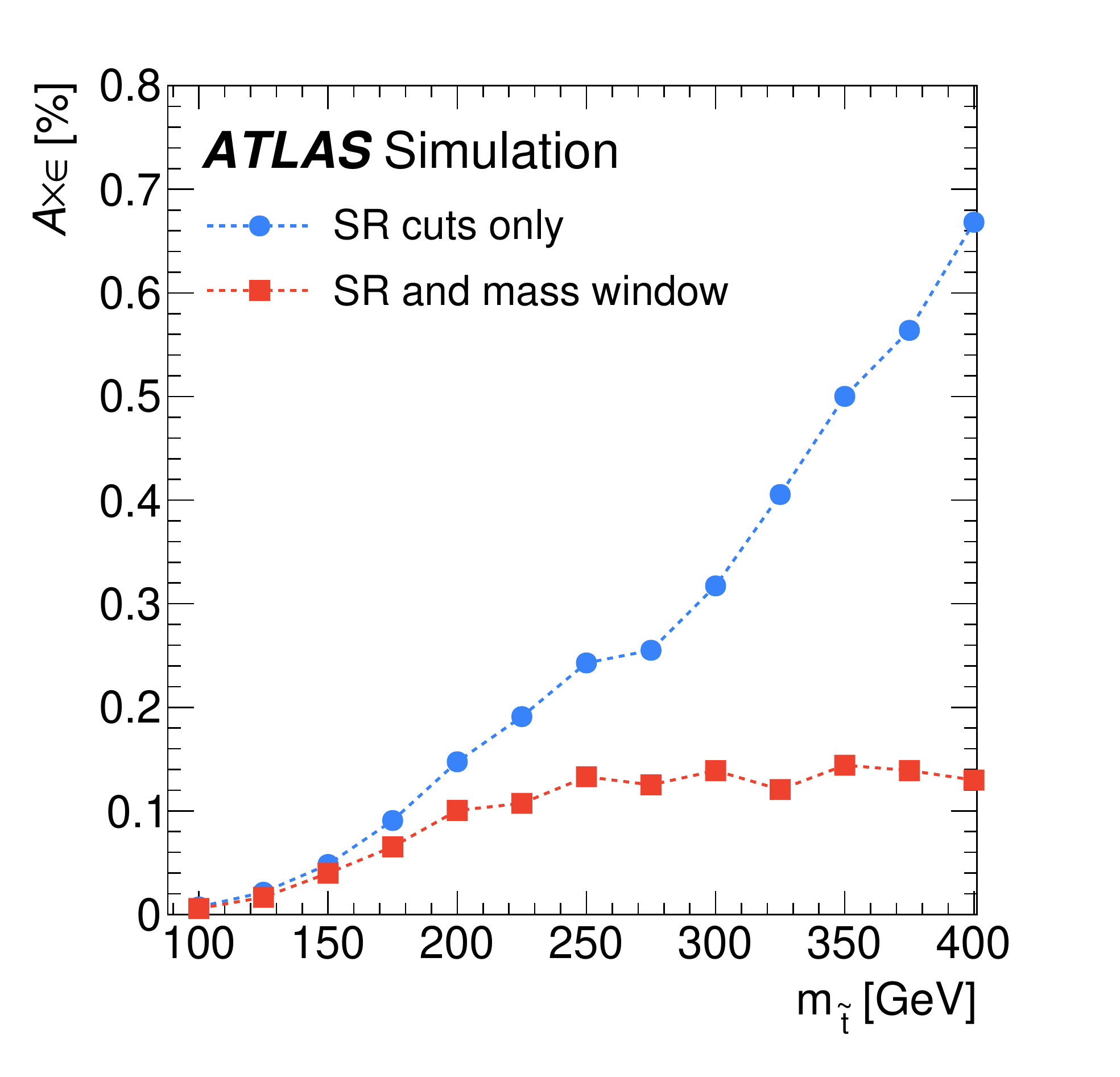}
    \caption{Total acceptance times efficiency ($A\times\epsilon$) of the SR cuts of \equref{SRcuts}, and SR cuts combined with the mass window selection in \tabref{masswindows:definition}, as a function of \mstop. The denominator of the efficiency (in \%) is the total number of events, i.e. the top row in \tabref{sig-control:signalcutflow}.}
    \label{fig:sigeffvsmass1D}
\end{figure}

Following these selections, the distribution of the average mass of the leading two \largeR \ jets, $\mavg = (m^{\rm jet}_1 + m^{\rm jet}_2) / 2$, is used to search for an excess of events above the background prediction. The search is done in regions of $m_{\mathrm{avg}}^{\mathrm{jet}}$ that are optimised to give the best significance. As shown in \figref{massres}, the stop signal is expected as a peak that would appear on top of a smoothly falling background spectrum. A Gaussian distribution is fitted to the stop signal \mavg \ peak. The mean of the fit, $\langle \mavg \rangle$, is consistent with \mstop in each case. The resolution of the \mavg \ peak is given approximately $s/\langle \mavg \rangle \sim 5-7\%$ (where $s$ is the standard deviation of the fit), and has only a weak dependence on the stop mass in the range probed by this analysis. Mass \textit{windows} in \mavg are determined by taking into account the effect of jet energy scale (JES) and jet energy resolution (JER) measurement uncertainties on the expected signal \mavg distribution and the estimated background. The size of each mass window is defined to be equal to or larger than the full width of the \mavg mass spectrum for the \mstop model that best corresponds to that range. The definitions of these mass windows and the signal efficiency in each window are given in \tabref{masswindows:definition}. \figref{massres:linear} shows the mass windows overlaid on top of the signal \mavg distributions for a few stop masses. The efficiency of the mass windows (relative to the SR cuts of \equref{SRcuts}) varies from $79\%$ at $100\gev$ to $19\%$ at $400\gev$. The low efficiency at high mass is due to the fact that the decay products are often not fully contained in the large-$R$ jet, as can be seen in \figref{massres:logy}. \figref{sigeffvsmass1D} shows the product of acceptance and efficiency, after the SR cuts and mass windows, as a function of \mstop. The significantly lower acceptance times efficiency for light stop masses in \figref{sigeffvsmass1D} is almost entirely due to the efficiency of the trigger selections which are for 100, 250, and 400 GeV stop masses 0.56\%, 22\%, and 96\%, respectively. This low efficiency is compensated by the large cross section for low stop masses retaining sensitivity to these mass values.

\begin{table}[!ht]
\footnotesize
\begin{center}\renewcommand\arraystretch{1.6}
\sisetup{round-mode=figures, round-precision=2,
retain-explicit-plus=true, group-digits = true}
\begin{tabular}{
c|c
S[table-format=2.1, table-number-alignment=center, round-mode=figures, round-precision=1]
}
\toprule
{$m_{\tilde{t}}\,[{\rm GeV}]$} & {Window [${\rm GeV}$]} & \multicolumn{1}{c}{Selection efficiency in mass window} \\
\midrule
$100$ & $[95, 115]$ & {\numRF{79.1775}{2}\,\%} \\
$125$ & $[115, 135]$ & {\numRF{77.4963}{2}\,\%} \\
$150$ & $[135, 165]$ & {\numRF{82.7224}{2}\,\%} \\
$175$ & $[165, 190]$ & {\numRF{72.0123}{2}\,\%} \\
$200$ & $[185, 210]$ & {\numRF{68.1666}{2}\,\%} \\
$225$ & $[210, 235]$ & {\numRF{56.1892}{2}\,\%} \\
$250$ & $[235, 265]$ & {\numRF{54.7513}{2}\,\%} \\
$275$ & $[260, 295]$ & {\numRF{49.0756}{2}\,\%} \\
$300$ & $[280, 315]$ & {\numRF{43.7924}{2}\,\%} \\
$325$ & $[305, 350]$ & {\numRF{29.7651}{2}\,\%} \\
$350$ & $[325, 370]$ & {\numRF{28.8176}{2}\,\%} \\
$375$ & $[345, 395]$ & {\numRF{24.6531}{2}\,\%} \\
$400$ & $[375, 420]$ & {\numRF{19.4232}{2}\,\%} \\
\bottomrule
\end{tabular}
\caption{Definition of the signal mass windows and selection efficiency in each window relative to the SR cuts of \equref{SRcuts}.}
\label{tab:masswindows:definition}
\end{center}
\end{table}


\newpage
\FloatBarrier

\section{Background estimation}
\label{sec:background}
The estimation of the dominant SM multijet background in the signal region, including both the expected number of events and the shape of the \mavg background spectrum, is performed directly from the data. MC simulations are used to study the background estimation method itself and to assess the contribution from \ttbar production. For the background estimation, additional kinematic regions are defined by inverting the \Asym \ and \costhstar \ selections as shown in \tabref{sig-control:SRCR}. These are labelled $An$, $Bn$, $Cn$, where $n$ indicates the number of $b$-tags ($\nbtag=0,\,=1,\,\geq2$). The signal region kinematic selection criteria of \equref{SRcuts} are comprised by the $Dn$ requirements and summarised in the last row of \tabref{sig-control:SRCR}, where $SR\equiv D2$ with $n\geq2$ $b$-tags, and $D1$ with $n=1$ $b$-tag is a validation region. 
Signal event yields are summarised in \tabref{sig-control:signalcutflow} for three stop masses.

\begin{table}[!ht]
\begin{center}\renewcommand\arraystretch{1.6}
\begin{tabular}{l |l |l |l | l}
\toprule
Region & {\Asym} & {\costhstar} & {\Ptratio} & \nbtag \\
\midrule
$An$ & $\geq 0.1$ & $\geq 0.3$ & $> 0.3$ & $ =0,\,=1,\,\geq2$ \\
$Bn$ & $< 0.1$    & $\geq 0.3$ & $> 0.3$ & $ =0,\,=1,\,\geq2$ \\
$Cn$ & $\geq 0.1$ & $< 0.3$    & $> 0.3$ & $ =0,\,=1,\,\geq2$ \\ 
\midrule
$Dn$ & $< 0.1$    & $< 0.3$    & $> 0.3$ & $ =0,\,=1,\,\geq2$ \\
\bottomrule
\end{tabular}
\caption{Definitions of the kinematic regions defined by \Asym, \costhstar, \ptratio, and the $b$-tag multiplicity ($\nbtag=0,\,=1,\,\geq2$). The letters $A$, $B$, $C$, and $D$ label the \Asym \ and \costhstar \ selections, whereas $\nbtag$ indicates the number of $b$-tags. $D2 \equiv SR$ is the signal region of the analysis.}
\label{tab:sig-control:SRCR}
\end{center}
\end{table}


\begin{table}[!ht]
\begin{center}\renewcommand\arraystretch{1.6}
\sisetup{round-mode=figures, round-precision=2,
retain-explicit-plus=true, group-digits = true}
\begin{tabular}{c | 
S
S
S
}
\toprule
Selection & {$\mstop = 100$~GeV} & {$\mstop = 250$~GeV} & {$\mstop = 400$~GeV} \\
\midrule
Total events & {$(9.72\pm0.01)\times10^6$} & {$(9.54 \pm 0.02)\times10^4$} & {$(6.202\pm0.002)\times10^3$} \\
Jet + $\HT$ trigger & {$(5.47\pm0.08)\times10^4$} & {$(2.07\pm0.01)\times10^4$} & {$(5.98\pm0.02)\times10^3$} \\
\LargeR jet tag & {$(1.68\pm0.04)\times10^4$} & {$(4.76\pm0.06)\times10^3$} & {$(1.29 \pm 0.01)\times10^3$} \\
$\nbtag \geq 2$ & {$(6.35\pm0.23)\times10^3$} & {$(1.70\pm0.03)\times10^3$} & {$515 \pm 6$} \\
\midrule
$A2$ & 416\pm58 & 194\pm11 & 68.7\pm2.2 \\
$B2$ & 639\pm71 & 199\pm11 & 33.3\pm1.6 \\
$C2$ & 419\pm62 & 149\pm9  & 71.2\pm2.2 \\
$D2$ & 711\pm74 & 240\pm12 & 41.5\pm1.8 \\
\bottomrule
\end{tabular}
\caption{The expected number of signal events in \intluminoerr from MC simulation for each of the selections applied to the $\nbtag \geq 2$ region. Stop masses of $\mstop = 100\GeV$, $250\GeV$ and $400\GeV$ are shown. The statistical uncertainty of the MC simulation is shown for each selection. The jet + \HT trigger selection includes the offline selection. The \largeR jet tag includes both the kinematic preselections and the \textit{splitting} criteria defined by \equref{splitting:asym} and \equref{splitting:mpt}. No selections are placed on the masses of the candidate stop jets. The region definitions of $A2$--$D2$ are summarised in \tabref{sig-control:SRCR}.}
\label{tab:sig-control:signalcutflow}
\end{center}
\end{table}

\begin{figure}[!ht]
    \centering
    \subfigure[]{
     \includegraphics[width=0.48\columnwidth]{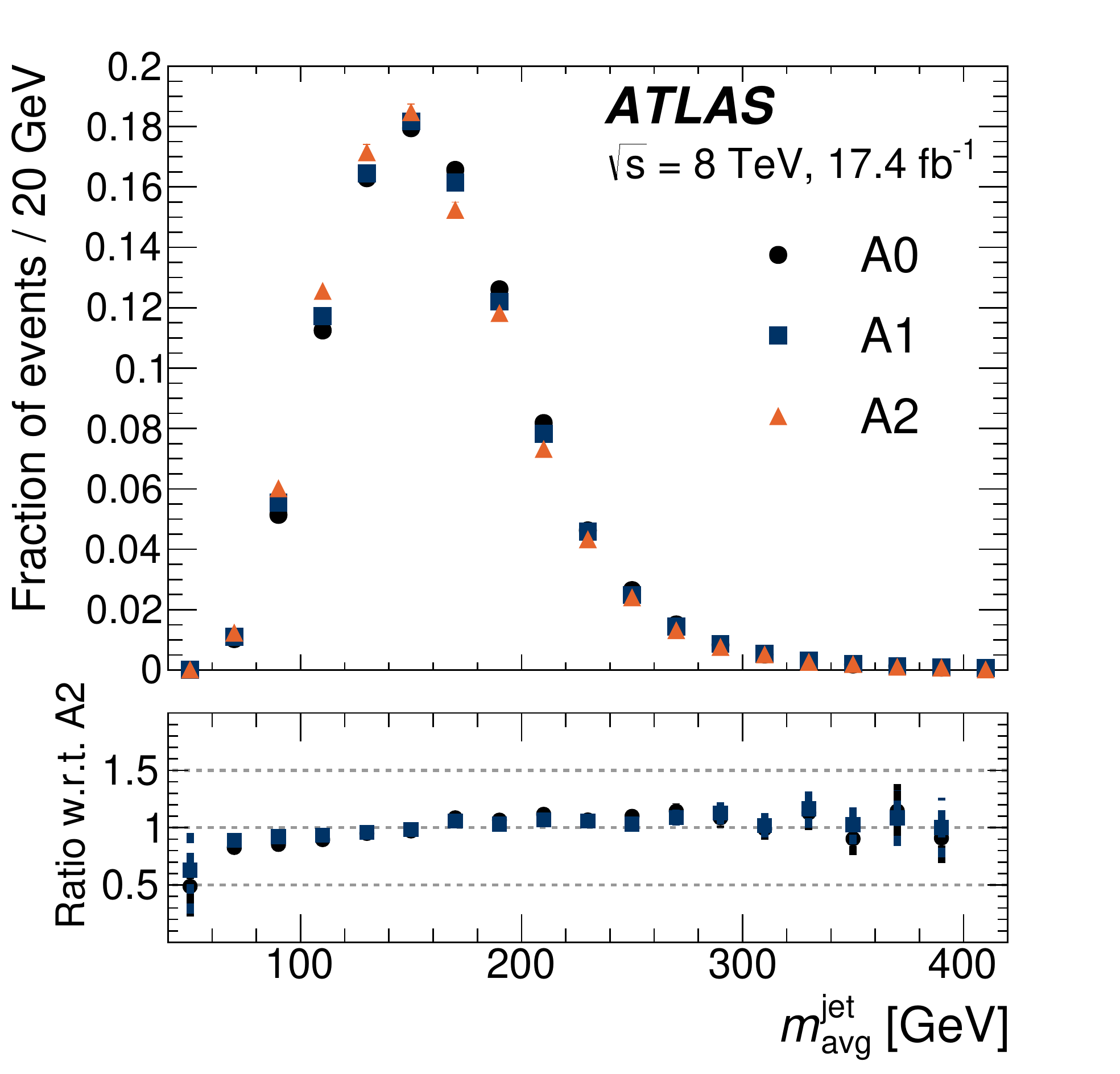}
      \label{fig:bckg:sidebands:regionA}
    }
    \subfigure[]{
     \includegraphics[width=0.48\columnwidth]{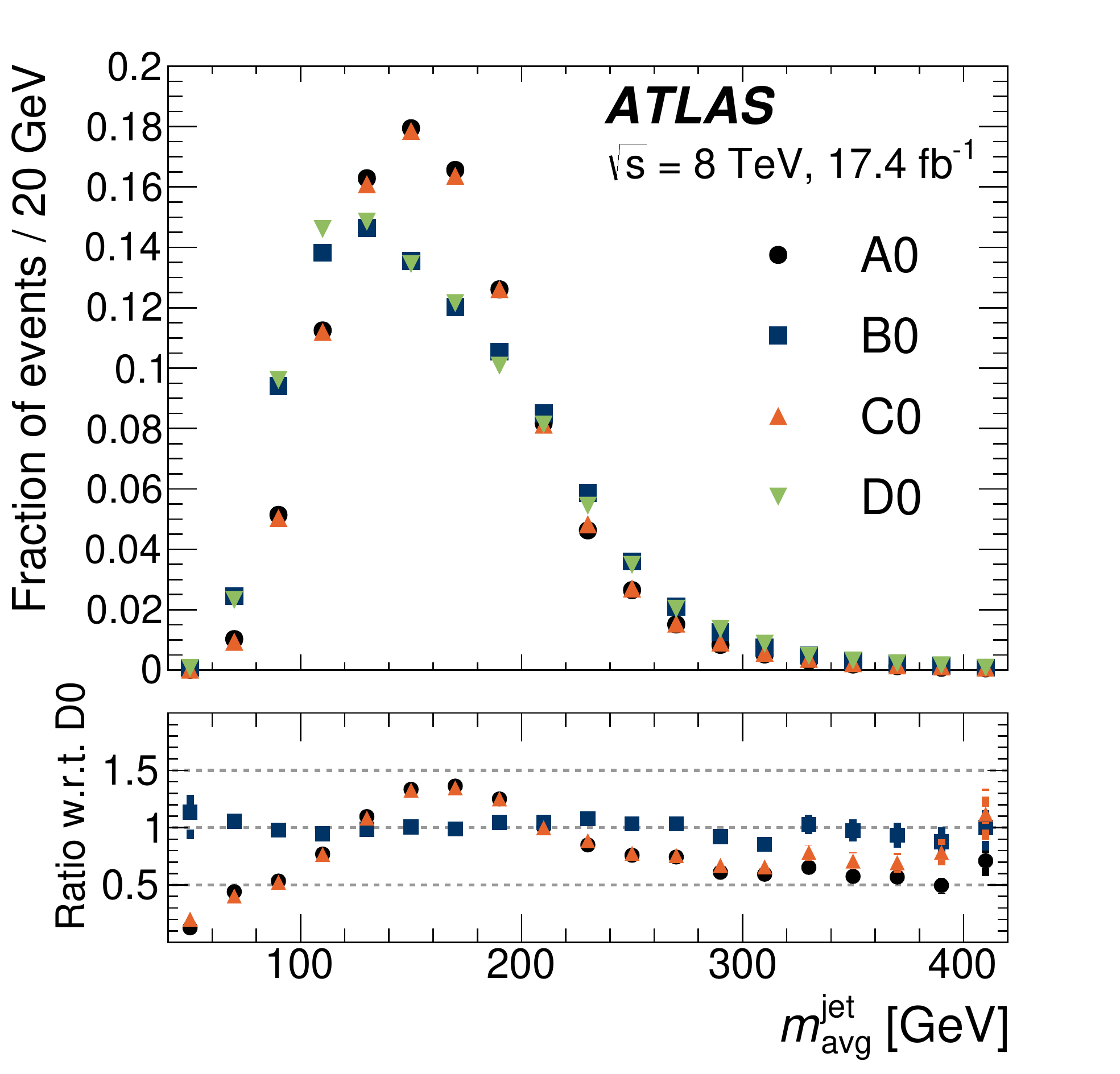}
      \label{fig:bckg:sidebands:nbtag0}
    }
    \caption{Shape comparisons of the \mavg spectrum for the data \subref{fig:bckg:sidebands:regionA} in region $A$ for events with $\nbtag=0,\,=1,\,\geq2$ and \subref{fig:bckg:sidebands:nbtag0} in regions $A,B,C,D$ for events with $\nbtag = 0$. In each case, the lower panel shows the ratio of the spectrum in each region to that which most closely matches the final signal region ($\nbtag \geq 2$ for region $A$ and region $D$ for $\nbtag = 0$). Only statistical uncertainties are shown. \label{fig:bckg:sidebands}}
\end{figure}

The method relies on the assumption that the shape of the \mavg spectrum is independent of the various $b$-tagging selections, as \figref{bckg:sidebands:regionA} indicates, in each of the kinematic regions ($An$, $Bn$, $Cn$, and $Dn$) defined in \tabref{sig-control:SRCR}. The advantage of the approach adopted here is that events with fewer than two $b$-tagged jets can be used as control and validation regions for \insitu studies of these kinematic regions. An estimation of the normalisation and shape of the spectrum in the signal region $D2$ can therefore be tested and validated using events with $\nbtag = 1$ as well as regions $A$ ($\Asym \geq 0.1$, $\costhstar \geq 0.3$) and $C$ ($\Asym \geq 0.1$, $\costhstar < 0.3$). Region $B$ ($\Asym < 0.1$, $\costhstar \geq 0.3$) is primarily used to evaluate shape differences in the predicted \mavg spectra (see \secref{bkgdest:shape}). 

The $\Asym$ and $\costhstar$ variables are found to have a correlation coefficient of at most $1\%$ in data events for $\nbtag = 0$. In simulated multijet events, the correlation is also consistent with zero in events with $\nbtag \geq 2$, within the large statistical uncertainties. Consequently, the ratio of $\nbtag \geq 2$ (or $\nbtag = 1$) to $\nbtag = 0$ in regions $A, B$, and $C$ should be approximately the same as the ratio in region $D$. The average jet mass spectrum, \mavg, is compared across the various $\nbtag$ selections for region $A$, as well as between each of the regions in events with $\nbtag = 0$. These comparisons are shown in \figref{bckg:sidebands} along with the ratio of the spectrum in each region to that which most closely matches the final signal region in each figure (region $D$ for $\nbtag = 0$ and $\nbtag \geq 2$ for region $A$). The results demonstrate that the \mavg spectra in regions $C$ and $D$ are reliably reproduced by regions $A$ and $B$, respectively, as shown in \figref{bckg:sidebands:nbtag0}. 

\begin{table}[!ht]
\footnotesize
\begin{center}\renewcommand\arraystretch{1.6}
\sisetup{round-mode=figures, round-precision=2,
retain-explicit-plus=true, group-digits = true}
\scalebox{0.90}{
\begin{tabular}{
c|
S[table-format=6.0, table-number-alignment=center, round-mode=places, round-precision=0]|
S[table-format=4.0, table-number-alignment=right, round-mode=places, round-precision=0]@{\quad$\pm\,$}
S[table-format=2.0, table-number-alignment=left, round-mode=places, round-precision=0]
S[table-format=2.0, table-number-alignment=left, round-mode=figures, round-precision=2]|
l r | l r | l r | l r 
}
\toprule
\multirow{2}{*}{Region} & \multirow{2}{*}{$N_{\rm data}$} & \multicolumn{3}{c|}{\multirow{2}{*}{$N_{\ttbar}$ ($\pm$ stat. $\pm$ syst.)}} & \multicolumn{2}{c|}{$[95, 115]\gev$} & \multicolumn{2}{c|}{$[135, 165]\gev$} & \multicolumn{2}{c|}{$[165, 190]\gev$} & \multicolumn{2}{c}{$[375, 420]\gev$} \\
& & \multicolumn{3}{c|}{} & $\frac{N_S}{N_{\rm data}}$ & $\frac{N_{\ttbar}}{N_{\rm data}}$ & $\frac{N_S}{N_{\rm data}}$ & $\frac{N_{\ttbar}}{N_{\rm data}}$ & $\frac{N_S}{N_{\rm data}}$ & $\frac{N_{\ttbar}}{N_{\rm data}}$ & $\frac{N_S}{N_{\rm data}}$& $\frac{N_{\ttbar}}{N_{\rm data}}$ \\
\midrule
\multicolumn{13}{c}{$\nbtag = 0$}\\
\midrule
$N_{A0}$ & 296226.00000 & {\phoo\phdo\numRF{388.62819}{2}} & 10.17313 & \numpmRF{+102.84027}{-94.90433}{2} & {\numRF{0.20767}{2}{$\,\%$}} & {\numRF{0.26805}{2}{$\,\%$}} & {\numRF{0.04840}{2}{$\,\%$}} & {\numRF{0.13575}{2}{$\,\%$}} & {\numRF{0.01911}{2}{$\,\%$}} & {\numRF{0.07188}{2}{$\,\%$}} & {\numRF{0.10929}{2}{$\,\%$}} & {\numRF{0.03717}{2}{$\,\%$}} \\
$N_{B0}$ & 115671.00000 & 175.68761 & 7.14731 & \numpmRF{+50.06439}{-41.90895}{2} & {\numRF{0.63587}{2}{$\,\%$}} & {\numRF{0.19547}{2}{$\,\%$}} & {\numRF{0.89621}{2}{$\,\%$}} & {\numRF{0.16597}{2}{$\,\%$}} & {\numRF{0.49503}{2}{$\,\%$}} & {\numRF{0.14263}{2}{$\,\%$}} & {\numRF{0.68376}{2}{$\,\%$}} & {\numRF{0.13476}{2}{$\,\%$}} \\
$N_{C0}$ & 114186.00000 & 220.93383 & 7.97008 & \numpmRF{+58.67726}{-52.01910}{2} & {\numRF{0.42460}{2}{$\,\%$}} & {\numRF{0.38832}{2}{$\,\%$}} & {\numRF{0.08841}{2}{$\,\%$}} & {\numRF{0.19834}{2}{$\,\%$}} & {\numRF{0.01955}{2}{$\,\%$}} & {\numRF{0.09338}{2}{$\,\%$}} & {\numRF{0.24312}{2}{$\,\%$}} & {\numRF{0.18195}{2}{$\,\%$}} \\
$N_{D0}$ & 44749.00000 & 110.47521 & 5.71055 & \numpmRF{+27.00959}{-27.10815}{2} & {\numRF{4.03156}{2}{$\,\%$}} & {\numRF{0.27317}{2}{$\,\%$}} & {\numRF{1.99419}{2}{$\,\%$}} & {\numRF{0.28690}{2}{$\,\%$}} & {\numRF{2.34189}{2}{$\,\%$}} & {\numRF{0.23538}{2}{$\,\%$}} & {\numRF{2.44933}{2}{$\,\%$}} & {\numRF{0.00000}{2}{$\,\%$}} \\
\midrule
\multicolumn{13}{c}{$\nbtag = 1$}\\
\midrule
$N_{A1}$ & 79604.00000 & {\phoo\numRF{1110.81971}{3}} & {\numRF{14.92183}{1}\pho} & \numpmRF{+189.22855}{-178.66436}{2} & {\numRF{1.24929}{2}{$\,\%$}} & {\numRF{2.55826}{2}{$\,\%$}} & {\numRF{0.46090}{2}{$\,\%$}} & {\numRF{1.48593}{2}{$\,\%$}} & {\numRF{0.48027}{2}{$\,\%$}} & {\numRF{0.73958}{2}{$\,\%$}} & {\numRF{0.21802}{2}{$\,\%$}} & {\numRF{0.70699}{2}{$\,\%$}} \\
$N_{B1}$ & 31045.00000 & 516.55041 & 10.63929 & \numpmRF{+83.98297}{-83.33829}{2} & {\numRF{14.19689}{2}{$\,\%$}} & {\numRF{1.88432}{2}{$\,\%$}} & {\numRF{9.73483}{2}{$\,\%$}} & {\numRF{2.27867}{2}{$\,\%$}} & {\numRF{7.97435}{2}{$\,\%$}} & {\numRF{1.94183}{2}{$\,\%$}} & {\numRF{10.06703}{2}{$\,\%$}} & {\numRF{0.08894}{2}{$\,\%$}} \\
$N_{C1}$ & 32163.00000 & {\phoo\phdo\numRF{624.58511}{2}} & {\numRF{11.32557}{1}\pho} & \numpmRF{+106.45048}{-100.81966}{2} & {\numRF{4.80324}{2}{$\,\%$}} & {\numRF{3.39138}{2}{$\,\%$}} & {\numRF{1.57510}{2}{$\,\%$}} & {\numRF{2.14990}{2}{$\,\%$}} & {\numRF{1.29479}{2}{$\,\%$}} & {\numRF{0.99450}{2}{$\,\%$}} & {\numRF{0.27908}{2}{$\,\%$}} & {\numRF{0.76052}{2}{$\,\%$}} \\
$N_{D1}$ & 12350.00000 & 306.15766 & 8.42288 & \numpmRF{+51.66637}{-44.96856}{2} & {\numRF{28.88994}{2}{$\,\%$}} & {\numRF{2.31882}{2}{$\,\%$}} & {\numRF{30.99461}{2}{$\,\%$}} & {\numRF{3.61402}{2}{$\,\%$}} & {\numRF{21.39476}{2}{$\,\%$}} & {\numRF{3.67838}{2}{$\,\%$}} & {\numRF{42.79505}{2}{$\,\%$}} & {\numRF{0.00010}{2}{$\,\%$}} \\
\midrule
\multicolumn{13}{c}{$\nbtag \geq 2$}\\
\midrule
$N_{A2}$ & 22259.00000 & {\phoo\numRF{1049.41011}{3}} & {\numRF{12.81141}{1}\pho} & \numpmRF{+191.87180}{-173.46193}{2} & {\numRF{2.24006}{2}{$\,\%$}} & {\numRF{6.80407}{2}{$\,\%$}} & {\numRF{1.72517}{2}{$\,\%$}} & {\numRF{5.74160}{2}{$\,\%$}} & {\numRF{1.15235}{2}{$\,\%$}} & {\numRF{2.75923}{2}{$\,\%$}} & {\numRF{1.01577}{2}{$\,\%$}} & {\numRF{1.89461}{2}{$\,\%$}} \\
$N_{B2}$ & 8416.00000 & 555.56676 & 9.79715 & \numpmRF{+94.07161}{-86.39268}{2} & {\numRF{49.75963}{2}{$\,\%$}} & {\numRF{7.23389}{2}{$\,\%$}} & {\numRF{29.22505}{2}{$\,\%$}} & {\numRF{9.98207}{2}{$\,\%$}} & {\numRF{24.08308}{2}{$\,\%$}} & {\numRF{8.75817}{2}{$\,\%$}} & {\numRF{26.37032}{2}{$\,\%$}} & {\numRF{0.24286}{2}{$\,\%$}} \\
$N_{C2}$ & 9384.00000 & {\phoo\phdo\numRF{573.52836}{2}} & {\numRP{10.}{0}\pho} & \numpmRF{+99.73013}{-93.52988}{2} & {\numRF{8.23901}{2}{$\,\%$}} & {\numRF{8.78245}{2}{$\,\%$}} & {\numRF{4.08858}{2}{$\,\%$}} & {\numRF{7.47421}{2}{$\,\%$}} & {\numRF{2.79320}{2}{$\,\%$}} & {\numRF{2.89518}{2}{$\,\%$}} & {\numRF{2.81337}{2}{$\,\%$}} & {\numRF{2.65917}{2}{$\,\%$}} \\
$N_{D2}$ & 3688.00000 & 311.08125 & 7.18802 & \numpmRF{+60.35603}{-47.34229}{2} & {\numRF{122.36083}{2}{$\,\%$}} & {\numRF{8.39661}{2}{$\,\%$}} & {\numRF{73.21088}{2}{$\,\%$}} & {\numRF{13.73374}{2}{$\,\%$}} & {\numRF{72.21593}{2}{$\,\%$}} & {\numRF{10.82466}{2}{$\,\%$}} & {\numRF{161.00079}{2}{$\,\%$}} & {\numRF{0.51163}{2}{$\,\%$}} \\
\bottomrule
\end{tabular}
}
\caption{The observed event yields for \intluminoerr in each of the regions for each $b$-tag multiplicity are shown, as well as the expected fractional signal contribution for the mass windows (as defined in \tabref{masswindows:definition}) corresponding to $\mstop=100$, 150, 175, and $400\gev$, and the \ttbar \ contribution in the same mass windows. The \ttbar \ systematic uncertainties include both the detector-level uncertainties and the theoretical uncertainties, as described in \secref{systematics}.  }
\label{tab:bkgdest:numbers}
\end{center}
\end{table}


The potential for events from \ttbar \ production to contribute increases with the addition of $b$-tag-multiplicity selections. \Tabref{bkgdest:numbers} presents the number of events in the data and the contribution from \ttbar, as determined by MC simulation, in regions $A$, $B$, $C$, and $D$ for $\nbtag=0,\,=1,\,\geq2$. The expected signal and \ttbar \ contributions are also given for a few mass windows. The \ttbar contribution is at the few per mille level in the events with $\nbtag = 0$. Contributions rise slightly in events with $\nbtag = 1$ to a maximum of $\lesssim 4\%$ in region $D1$. Lastly, regions $A2$ and $C2$ ($\Asym \geq 0.1$) have a maximum \ttbar \ contribution of around $\lesssim10\%$. Consequently, when validating the method and in the final background estimate, the contribution from \ttbar \ is subtracted in each of the regions. The corrected total number of events in a given region is defined as $N_{Xn} = N_{Xn}^{\rm data} - N_{Xn}^{\ttbar}$ and the corrected \mavg \ spectrum is defined as $N_{Xn,i} = N_{Xn,i}^{\rm data} - N_{Xn,i}^{\ttbar}$, where $i$ represents the $i^{\rm th}$ bin of the histogram ($X=A,B,C,$ or $D$, and $n$ refers to the number of $b$-tags). The two quantities are related by $N_{Xn} = \Sigma_i N_{Xn,i}$.

All regions used for the background estimation ($A0$, $C0$, $D0$, $A2$, and $C2$) exhibit potential signal contribution of less than $10\%$. Region $B2$ ($\Asym<0.1$, $\costhstar \geq 0.3$) is not used to derive the background estimate, since the expected signal contribution is much higher here than in $A2$ and $C2$ (for $\mstop=100\gev$ the signal contribution is $50\%$ in $B2$, compared with $2.2\%$ in $A2$ and $8.2\%$ in $C2$). The expected signal contribution in the validation regions ($\nbtag=1$) is only significant in $B1$ and $D1$ (both require $\Asym < 0.1$). Due to this level of expected signal contribution, and the \mavg dependence of that contribution, the background estimation procedure obtains the \mavg spectrum from the \nbtag = 0 regions for the final background spectrum estimate. The background estimation procedure itself is summarised in the following steps: 
\begin{enumerate}
  \item The \mavg \ shape ($N_{D0,i}$) and total number of events ($N_{D0}$) are extracted from the $D0$ region.
  \item A projection factor is derived between events with $\nbtag = 0$ and events with $\nbtag \geq 2$ for the signal-depleted regions $A$ ($\Asym \geq 0.1$, $\costhstar \geq 0.3$) and $C$ ($\Asym \geq 0.1$, $\costhstar < 0.3$). As explained above, the number of \ttbar events is subtracted in regions $A0, C0, A2,$ and $C2$ before evaluating the projection factor $\kavg$:
    \begin{equation}
        \kavg = (k_{A2} + k_{C2})/2,\quad \mbox{where} \quad k_{X2} = \frac{N_{X2}}{N_{X0}},\quad X=A,C.
        \label{eq:projfactor:avg:ttbar}
    \end{equation}
    \item The projection factor is used to estimate the total number of events,
        \begin{equation}
            N_{D2}^{'} = \kavg \times N_{D0} + N^{\ttbar}_{D2},
        \end{equation}
        and shape (bin-by-bin),
        \begin{equation}
          N_{D2,i}' = \kavg \times N_{D0,i} + N^{\ttbar}_{D2,i},
        \end{equation}
    in the signal region, $D2$ (where the contribution from \ttbar \ in $D2$ has been added).
\end{enumerate}
This procedure is performed in the entire mass range and the mass windows are then defined from the estimated background spectrum. 
The projection factors $k_{A2}$ and $k_{C2}$ are compatible at the level of about $4\%$ (including the \ttbar \ subtraction as in \equref{projfactor:avg:ttbar}) and this difference is included as a systematic uncertainty on the background estimate (see \secref{systematics}). The validity of the background estimation method can be demonstrated in the $\nbtag = 1$ regions by deriving a projection factor analogously to \equref{projfactor:avg:ttbar} for $\nbtag = 0$ and $\nbtag = 1$, 
\begin{equation}
\kavgvalid = (k_{A1} + k_{C1})/2.
\label{eq:projfactor:avg:valid}
\end{equation}
The expected number of events in the full range of $D1$ is then estimated by 
\begin{eqnarray}
N_{D1}^{\prime} &=& \kavgvalid\times N_{D0} + N^{\ttbar}_{D1} \nonumber \\
                &=& 12400\pm130.
\label{eq:nonclosure:ND1}
\end{eqnarray}
The same estimate for $D2$ gives 
\begin{eqnarray}
N_{D2}' &=& \kavg\times N_{D0} + N^{\ttbar}_{D2} \nonumber \\
        &=& 3640^{+90}_{-80}.
\label{eq:nonclosure:ND2}
\end{eqnarray}
In \equref{nonclosure:ND1} and \equref{nonclosure:ND2} the uncertainty quoted includes the statistical uncertainty and the uncertainties related to the \ttbar \ estimate (see \secref{systematics}). These numbers should be compared with the observed numbers of events in \tabref{bkgdest:numbers}, $12350$ in $D1$ and $3688$ in $D2$. The observed numbers of events are  consistent with the estimated values.

\section{Systematic uncertainties}
\label{sec:systematics}
Several sources of systematic uncertainty are considered when determining the estimated contributions from signal and background. The background estimate uncertainties pertain primarily to the method itself. The control and validation regions defined in \secref{background} are used to evaluate the size of these uncertainties. A description of the primary sources of uncertainty follows.

\subsection{$b$-jet-multiplicity \mavg \ shape uncertainty}
\label{sec:bkgdest:btag}
Regions $A$ ($\Asym \geq 0.1$, $\costhstar \geq 0.3$) and $C$ ($\Asym \geq 0.1$, $\costhstar < 0.3$)  are used to directly compare the shape of the \mavg spectrum in events with $b$-jet-multiplicities of $\nbtag = 0$ and $\nbtag\geq2$ (the \ttbar-corrected \mavg spectrum is used, as defined in \secref{background}). The $b$-jet-multiplicity \mavg \ shape systematic uncertainty is calculated as the maximum of the bin-by-bin difference of region $A2$ compared to $A0$ (\figref{bckg:sidebands:regionA}) and $C2$ compared to $C0$,
\begin{equation}
  \sigma_i^{b{\rm -jet-multi.~syst.}} = {\rm \max}\left[|1-\nu_{A2,i}/\nu_{A0,i}|, |1-\nu_{C2,i}/\nu_{C0,i}|\right],
  \label{eq:bjetshapesyst}
\end{equation}
where the normalised \mavg \ spectrum are defined as $\nu_{Xn,i}=N_{Xn,i}/N_{Xn}$ ($X=A,C$).
The expression in \equref{bjetshapesyst} is then added in quadrature with the statistical uncertainty to form the total systematic uncertainty for that particular bin. A fixed bin width of 50 GeV is used in order to reduce effects due to statistical uncertainties.
The size of the $b$-jet-multiplicity \mavg \ shape systematic uncertainty varies from approximately 7--12$\%$ at low \mavg to 20\% near $\mavg\approx300\GeV$, and to around 90\% for $\mavg\approx400\GeV$. The large systematic uncertainty in the high-mass tail is due to the low number of events in the $\nbtag\geq2$ regions. 
\Figref{syst:totalbckg} shows the $b$-jet-multiplicity \mavg \ shape systematic uncertainty as well as the total systematic uncertainty when combined with the constant systematic uncertainty due to the $4\%$ difference between projection factors $k_{A2}$ and $k_{C2}$ mentioned in \secref{background}, and the background estimation \mavg \ shape systematic uncertainty described below in \secref{bkgdest:shape}.

\subsection{Background estimation \mavg \ shape uncertainty}
\label{sec:bkgdest:shape}
\begin{figure}[!ht]
  \centering
  \subfigure[Region A]{
    \includegraphics[width=0.47\columnwidth]{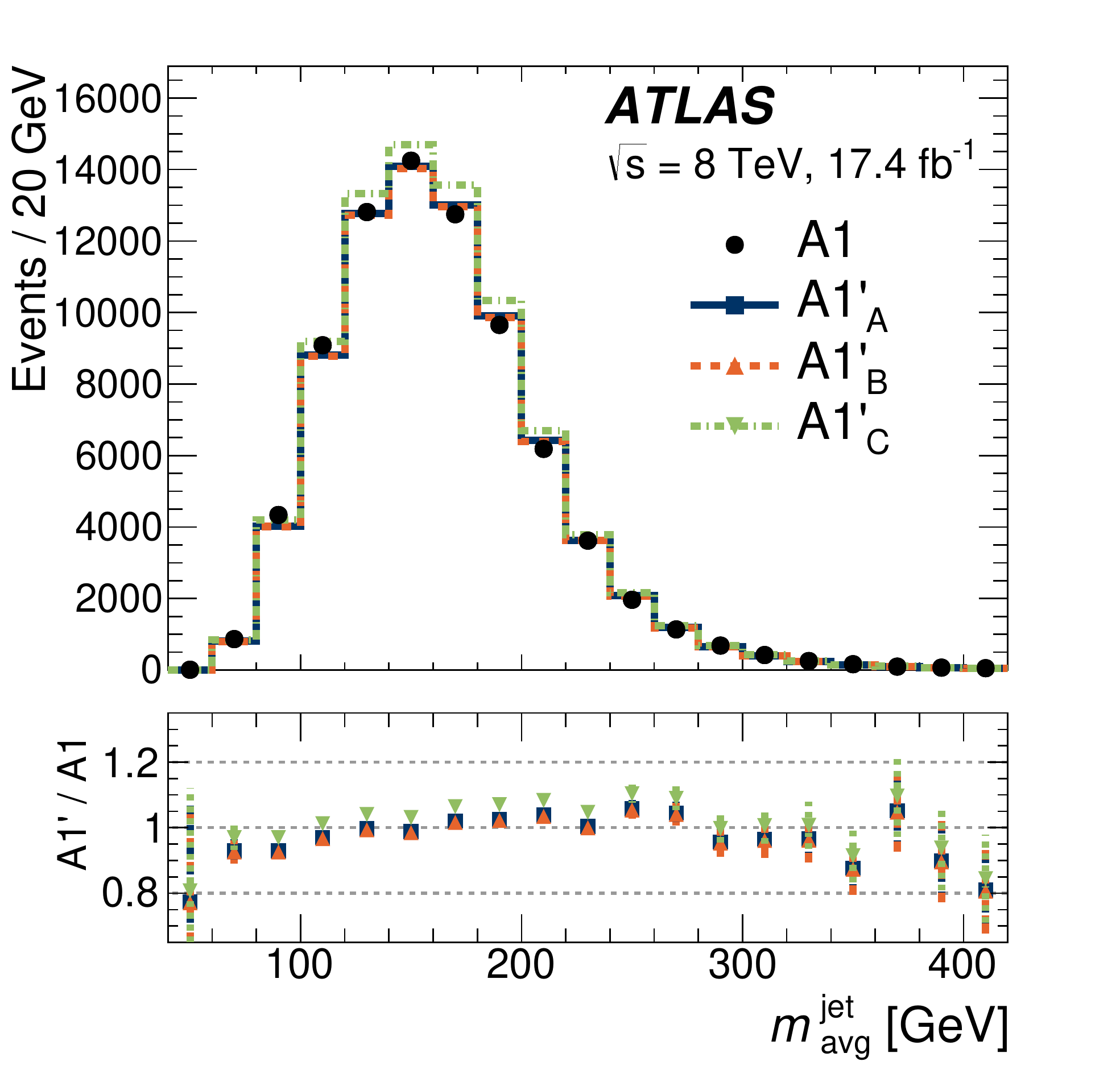}
    \label{fig:bckg:closure:validation:vr1:data:SB-A}
  }
  \subfigure[Region B]{
    \includegraphics[width=0.47\columnwidth]{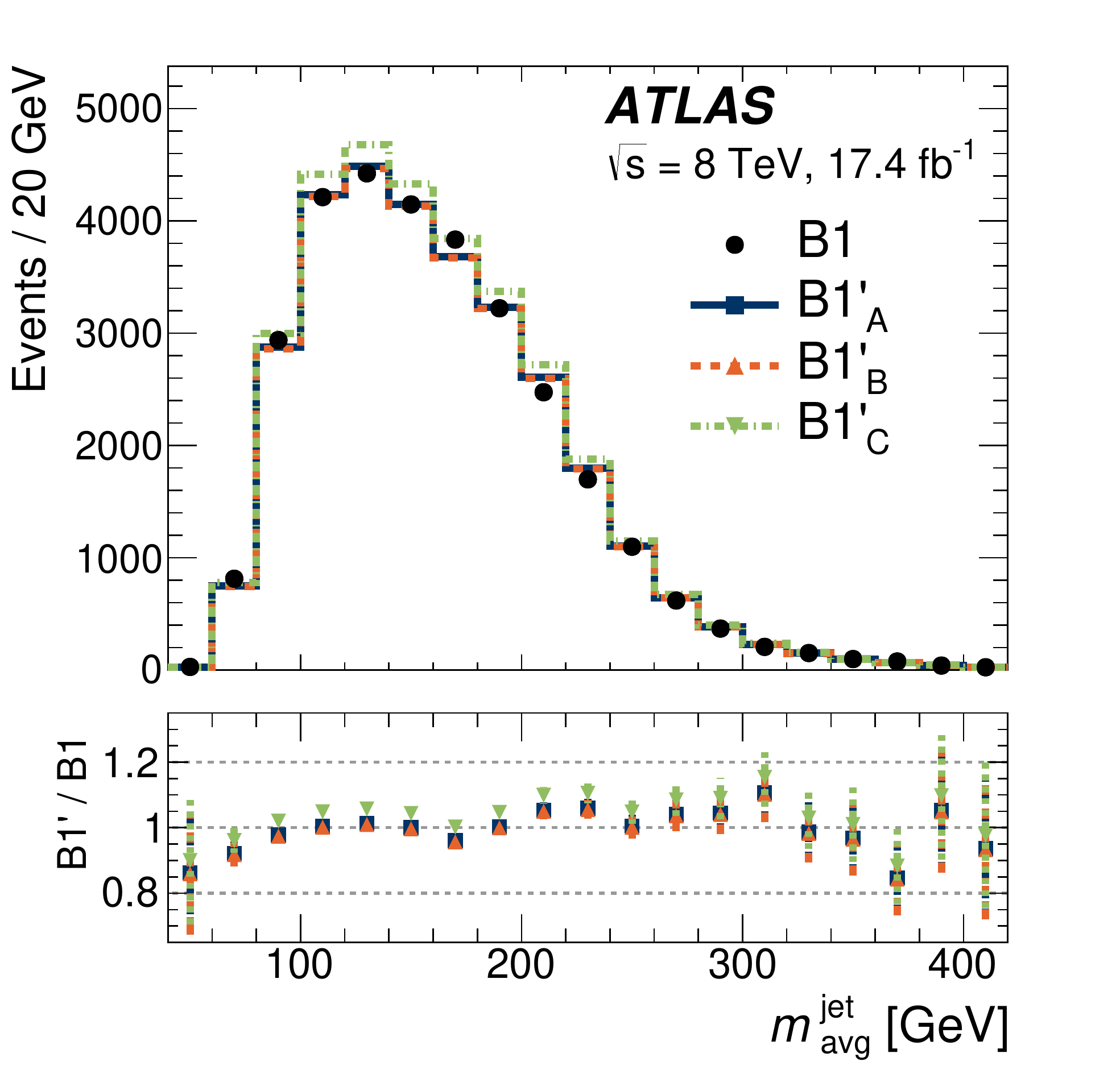}
    \label{fig:bckg:closure:validation:vr1:data:SB-B}
  } \\
    \subfigure[Region C]{
      \includegraphics[width=0.47\columnwidth]{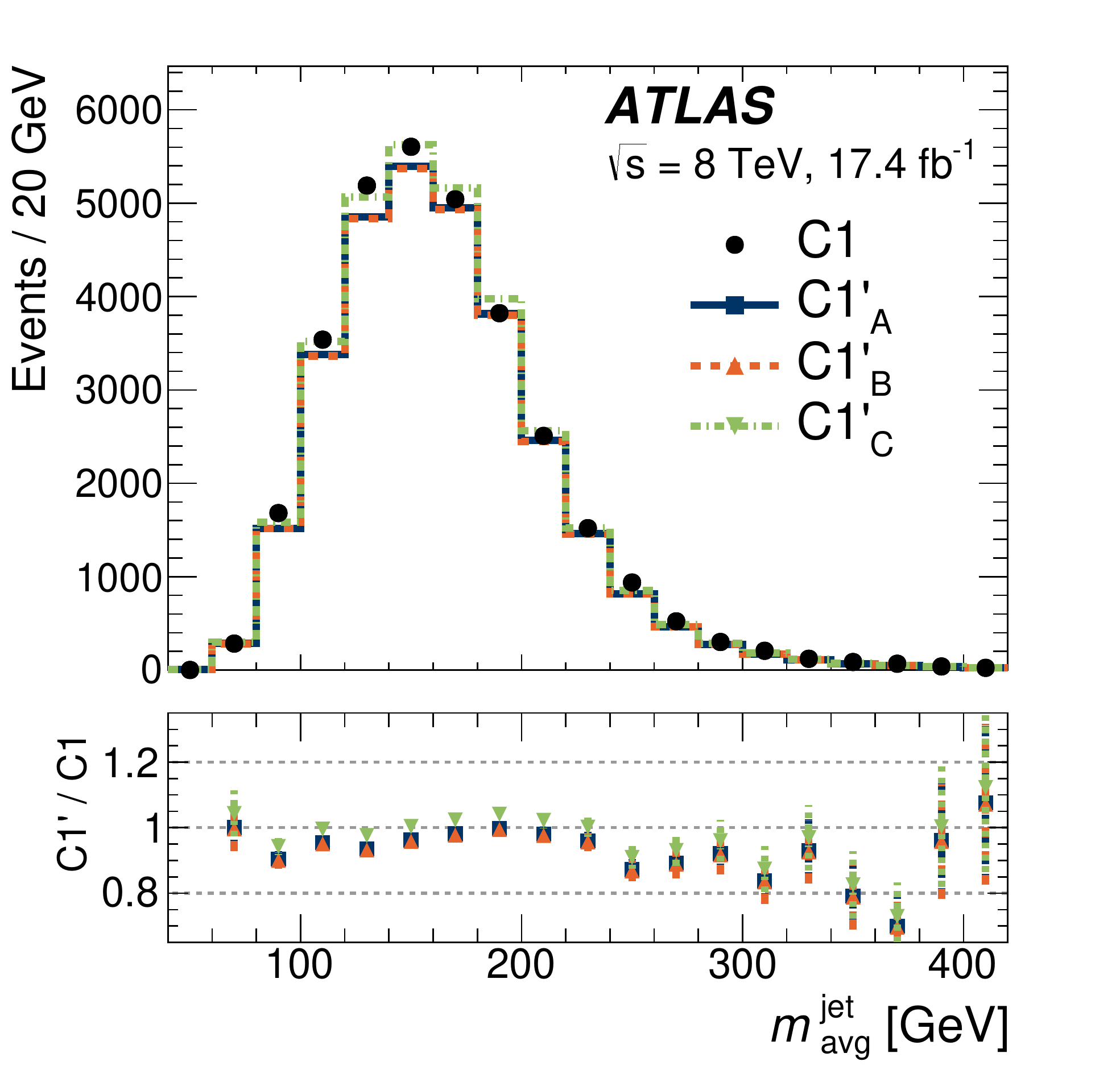}
      \label{fig:bckg:closure:validation:vr1:data:SB-C}
  }
    \subfigure[Region D]{
      \includegraphics[width=0.47\columnwidth]{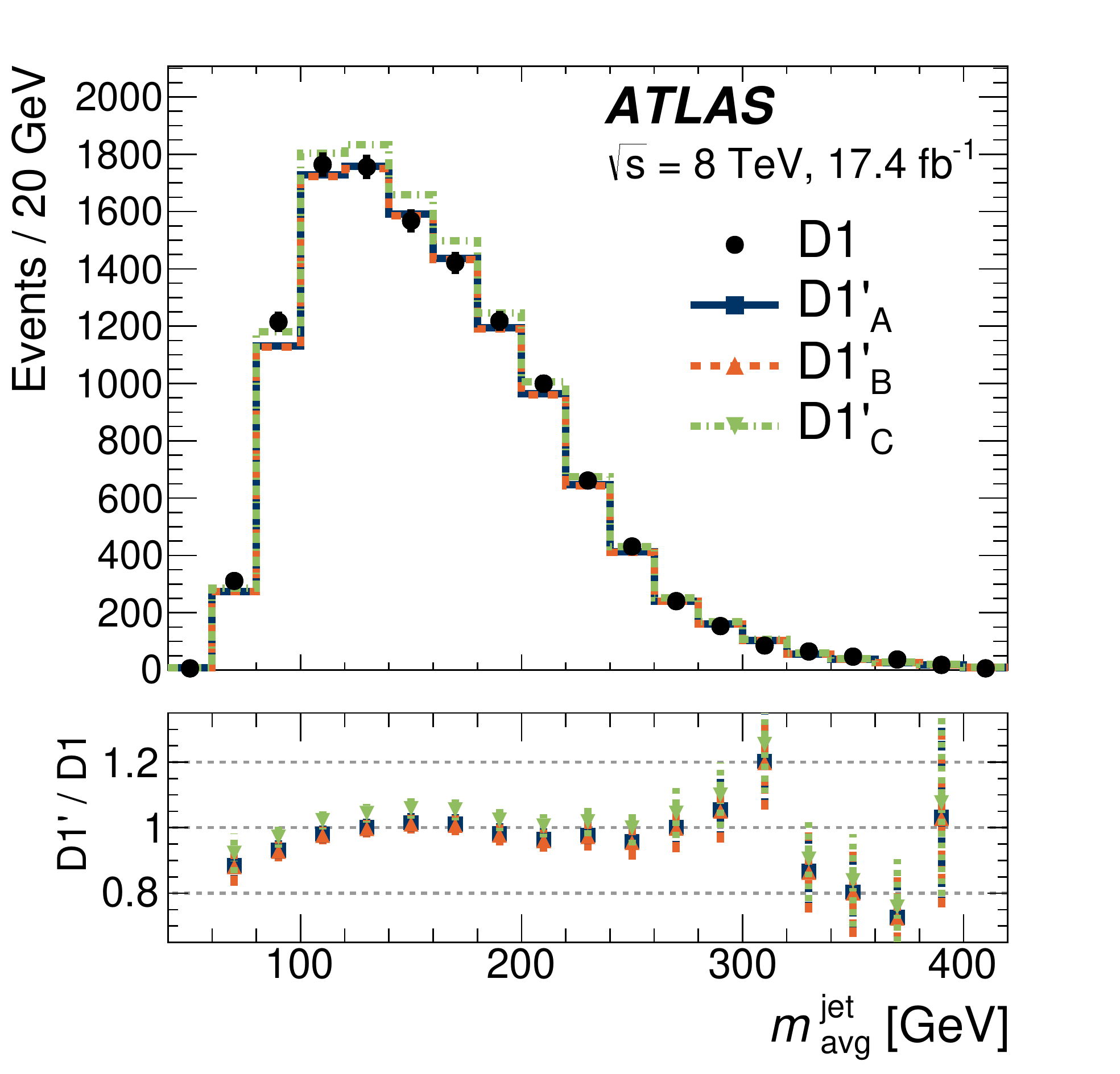}
      \label{fig:bckg:closure:validation:vr1:data:SB-D}
  }~
  \caption{The \mavg \ distribution is shown in four validation regions with $\nbtag = 1$. In each case the data ($A1$, $B1$, $C1$, and $D1$) are compared to estimates based on projection factors derived between $\nbtag=0$ and $\nbtag=1$ in $A$, $B$, and $C$ (see \secref{bkgdest:shape}).
  \label{fig:bckg:closure:validation:vr1:data}}
\end{figure}

\begin{figure}[!ht]
    \centering
    \includegraphics[width=0.75\columnwidth]{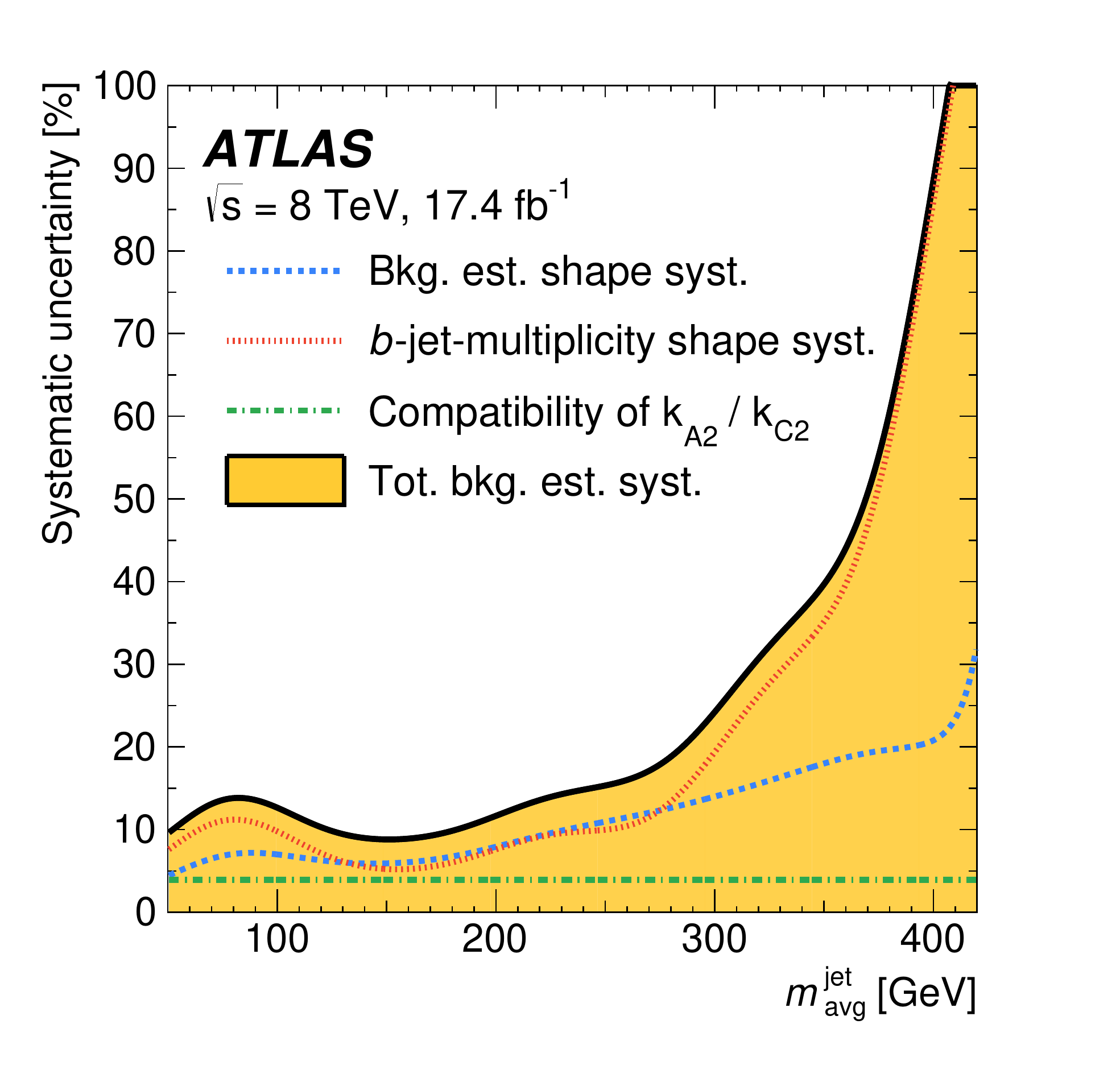}
    \caption{Systematic uncertainty for the data-driven multijet background estimation. The blue dashed line represents the background estimation systematic uncertainty estimated from comparisons of the predicted \mavg spectra in regions $A1$, $B1$, and $C1$ to the actual spectra. The red dotted line represents the estimated systematic uncertainty due to shape differences between events with $\nbtag = 0$ and $\nbtag \geq 2$. The green line represents a systematic uncertainty due to the level of compatibility of $k_{A2}$ and $k_{C2}$. Finally, the black line with a filled yellow area shows the combined systematic uncertainty of all three contributions added in quadrature. The systematic uncertainty curves were smoothed with a Gaussian filter of spread $20\gev$.}
 \label{fig:syst:totalbckg}
\end{figure}

Events with $\nbtag = 1$ are used to test the validity of the background estimation method in data and to derive a systematic uncertainty on the approach. \figref{bckg:closure:validation:vr1:data} shows several results of this test by comparing three estimated spectra with the observed spectrum in each of the four regions. The estimated spectra of \figref{bckg:closure:validation:vr1:data} are determined using projection factors, 
\begin{equation}
  k_{X1} = N_{X1}/N_{X0}, 
  \label{eq:validation:kX1}
\end{equation}
from events with $\nbtag = 0$ to those with $\nbtag = 1$, in each of the three regions $X=A$, $B$, and $C$ in order to determine the extent to which the prediction varies with each choice. Region $D1$ was used to validate the systematic uncertainty derived from $A1$, $B1$, and $C1$. Because of the three projection factors ($k_A$, $k_B$, and $k_C$) there are three estimates ($N_{Y1'_{A},i}$, $N_{Y1'_{B},i}$, and $N_{Y1'_{C},i}$) of the \mavg \ spectrum in each of the regions $Y1=A1$, $B1$, and $C1$. Thus, in total there are nine estimates of the actual spectra, these are written succinctly as
\begin{equation}
  N_{Y1'_{X},i} = k_{X1}\times N_{Y0,i} \mbox{, where $X=\{A,B,C\}$ and $Y=\{A,B,C\}$}.
  \label{eq:validation:vr1b}
\end{equation}
These estimates provide a test of the shape compatibility as well as the overall normalisation of the background estimate (the special cases $N_{A1'_{A},i}$, $N_{B1'_{B},i}$, and $N_{C1'_{C},i}$ are normalised to the data by construction and thus only provide a shape comparison of $\nbtag=1$ and $\nbtag=0$). A systematic uncertainty for the background projection is then derived by taking, bin-by-bin, the largest deviation of the ratio of estimated to actual yield from unity in the \mavg \ spectra in each of the regions $A$, $B$, and $C$ according to
\begin{equation}
  \sigma^{\rm bkg.~syst.}_i =  
  {\rm \max\limits_{X,Y}}\left[|1-N_{Y1'_{X},i} / N_{Y1,i}|\right],
\end{equation}
where $N_{Y1,i}$ are the observed data points and $N_{Y1'_{X},i}$ are the estimated spectra defined by \equref{validation:vr1b}. A bin width of 50 GeV is used, just as above with the $b$-jet multiplicity \mavg \ shape systematic uncertainty. This is added in quadrature with the statistical uncertainty of that ratio in order to form the total systematic uncertainty for that particular bin. 
The size of the background estimation \mavg \ shape systematic uncertainty varies from less than 10\% at low $\mavg\approx100\GeV$ to 20\% near $\mavg\approx400\GeV$. \Figref{syst:totalbckg} shows the background estimation \mavg \ shape systematic uncertainty as well as the total systematic uncertainty when combined with the two above-mentioned systematic uncertainties.

\subsection{Background \ttbar contribution systematic uncertainty}
Since \PowPythia MC simulation is used to determine the contribution from \ttbar events in the signal region and each of the control regions, systematic uncertainties related to the MC simulation of the process itself are included in the total systematic uncertainty for the background estimation. 
The theoretical uncertainties include renormalisation and factorisation scale variations, parton distribution function uncertainties, the choice of MC generator using comparisons with \texttt{MC@NLO}~\cite{Frixione:2002ik}, the choice of parton shower models using comparisons with \Herwig~\cite{Corcella:2000bw}, and initial- and final-state radiation (FSR) modelling uncertainties. The size of the theoretical systematic uncertainties for \ttbar production vary from approximately 40\% to 70\% in the relevant kinematic regions and are dominated by the uncertainties from the MC generator and ISR/FSR variations.
The detector-level uncertainties include the JES and JER uncertainties~\cite{PERF-2012-01} as well as the $b$-tagging efficiency and mistag-rate uncertainties~\cite{Aad:2015ydr}. Uncertainties associated with the \largeR jet mass scale and resolution are taken into account by the JES and JER uncertainties of the input small-$R$ jets~\cite{Nachman:2014kla}.

The size of the total \ttbar \ systematic uncertainty varies in the mass range $\mavg=100$--$200\gev$ from approximately $50\%$ to $80\%$. In the range $\mavg=300$--$400\gev$ the \ttbar \ systematic uncertainties are of the order of $100\%$, but the \ttbar background is completely negligible in this range. 
Lastly, an uncertainty of 2.8\% is applied to the measured integrated luminosity of \intluminoerr following the methodology described in Ref.~\cite{Aad:2013ucp}.

\subsection{Signal systematic uncertainties}
In addition to the systematic uncertainties associated with the background estimate, the MC simulation of the signal model is subject to systematic uncertainties. Much like the contribution from \ttbar, these uncertainties include experimental uncertainties as well as theoretical uncertainties. The  detector-level uncertainties include the JES and JER uncertainties, and the $b$-tagging uncertainties as described for the estimate of \ttbar. The theoretical uncertainties include renormalisation and factorisation scale variations, parton distribution function uncertainties, and ISR and FSR modelling uncertainties. The nominal signal cross-section and its uncertainty are taken from an envelope of cross-section predictions using different PDF sets and factorisation and renormalisation scales, as described in Ref.~\cite{Kramer:2012bx}. Each signal model is varied according to these systematic uncertainties and the impact on the acceptance in each mass window is then propagated to the final result. 
The largest contribution to the total signal systematic uncertainty comes from the JES and $b$-tagging, both in the range 10--18$\%$. 
The size of the  theoretical uncertainty grows from around $5\%$ for low-mass stops to around $10\%$ for higher-mass stops. 

To evaluate the ISR/FSR systematic uncertainty, separate samples of $\stop\stop^*$ pair events are generated using \Madgraph+\Pythia, and the rate of ISR/FSR production is varied. These are used to reweight the $\pt(\stop\stop^*)$ distribution of the nominal signal samples to estimate the change in signal acceptance $\times$ efficiency. The effect ranges from 0--17$\%$, with the largest impact at high \mstop.

\section{Results}
\label{sec:results}
\begin{figure}[!ht]
    \centering
    \includegraphics[width=0.85\columnwidth]{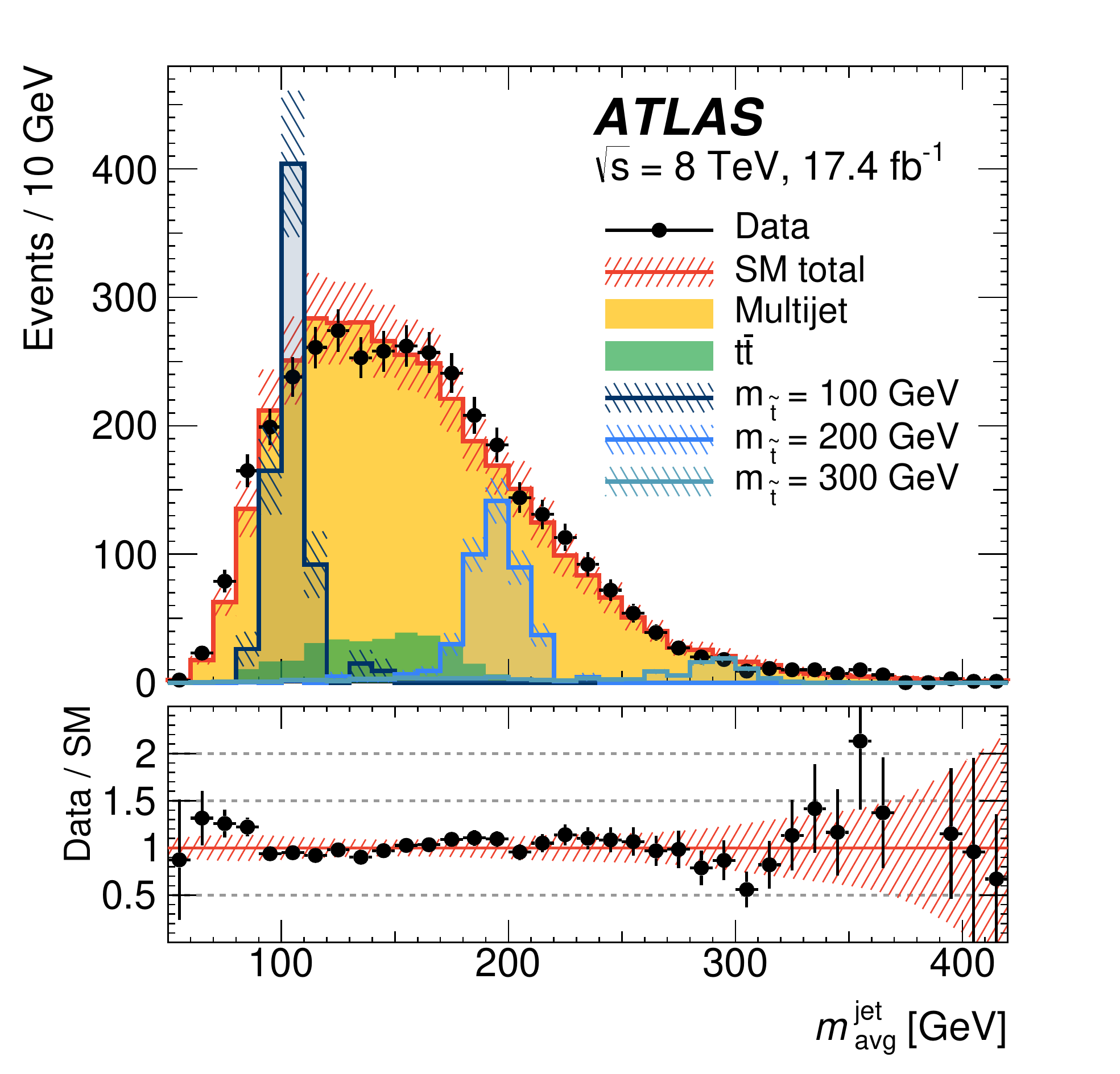}
    \caption{The observed \mavg \ spectrum in the signal region is shown as black points with statistical uncertainties. Also shown is the total SM background estimate, and the separate contributions from the data-driven multijet and MC \ttbar backgrounds. The red hatched band represents the combined statistical and systematic uncertainty on the total SM background estimate. Signal mass spectra are shown with statistical uncertainties only. The bottom panel shows the ratio of the data relative to the total SM background estimate.}
  \label{fig:bckg:est-shape}
\end{figure}

\tabref{results:windows} summarises the observed and expected number of events that fall within each of the optimised mass windows in the signal region, $D2$. \figref{bckg:est-shape} shows the observed \mavg distribution in the data, along with the estimated background spectrum, including both the systematic and statistical uncertainties. No excess over the background prediction is observed.

\begin{table}[!ht]
\footnotesize
\begin{center}\renewcommand\arraystretch{1.6}
\sisetup{round-mode=figures, round-precision=2,
retain-explicit-plus=true, group-digits = true}
\begin{tabular}{
c | c|
S[table-format=4.1, table-number-alignment=center, round-mode=places, round-precision=0] @{$\,\pm\,$}
S[table-number-alignment=center, round-mode=figures, round-precision=2] |
S[table-format=2.4, table-number-alignment=center, round-mode=figures, round-precision=2] @{$\,\pm\,$}
S[table-number-alignment=center, round-mode=figures, round-precision=2] |
S[table-format=3.1, table-number-alignment=center, round-mode=places, round-precision=0] @{$\,\pm\,$}
S[table-number-alignment=center, round-mode=figures, round-precision=2] |
S[table-format=3.0, table-number-alignment=center, round-mode=places, round-precision=0]|
S[table-format=3.1, table-number-alignment=center, round-mode=figures, round-precision=2] @{$\,\pm\,$}
S[table-number-alignment=center, round-mode=figures, round-precision=2]
}
\toprule
{$m_{\tilde{t}}\,[{\rm GeV}]$} & {Window [${\rm GeV}$]} & \multicolumn{2}{c|}{$N_{B}^{\rm data\mbox{-}driven~est.}$} & \multicolumn{2}{c|}{\quad $N_{B}^{\ttbar~\rm est.}$ \quad} & \multicolumn{2}{c|}{\quad $N_{B}^{\rm tot.~est.}$ \quad} & {\quad $N_{\rm data}^{\rm obs.}$ \quad} & \multicolumn{2}{c}{\quad $N_S$ \quad}  \\
\midrule
$100$ & $[95, 115]$ & 465.1841 & 55.5347 & 38.6244 & 26.0351 & 503.8085 & 61.3346 & 460.0000 & 562.8598 & 144.1721 \\
$125$ & $[115, 135]$ & 496.1020 & 49.0525 & 67.6184 & 36.9047 & 563.7204 & 61.3849 & 555.0000 & 570.8616 & 127.3171 \\
$150$ & $[135, 165]$ & 680.1193 & 60.9960 & {\numRP{104.5138}{0}\phoo\phoo\phdo} & 49.0779 & 784.6331 & 78.2889 & 761.0000 & 557.1348 & 111.3102 \\
$175$ & $[165, 190]$ & 470.9951 & 46.4012 & 63.1078 & 19.3233 & 534.1029 & 50.2639 & 583.0000 & {\numRP{421.0189}{0}\pho} & 96.2130 \\
$200$ & $[185, 210]$ & 395.1531 & 45.6549 & {\numRP{16.4775}{1}\pho\phoo} & 9.5626 & 411.6306 & 46.6456 & 416.0000 & {\numRP{293.4572}{0}\pho} & 49.9941 \\
$225$ & $[210, 235]$ & 266.4155 & 36.8115 & 2.4153 & 2.4153 & 268.8307 & 36.9033 & 283.0000 & {\numRP{177.6736}{0}\pho} & 35.5366 \\
$250$ & $[235, 265]$ & 175.8223 & 27.2307 & 1.1009 & 1.1009 & 176.9232 & 27.2723 & 195.0000 & {\numRP{126.8421}{0}\pho} & 29.3757 \\
$275$ & $[260, 295]$ & 103.7799 & 19.3482 & 0.5927 & 0.5495 & 104.3726 & 19.3561 & 96.0000 & 71.3053 & 20.2511 \\
$300$ & $[280, 315]$ & 68.6899 & 16.1502 & 0.9333 & 0.2887 & 69.6232 & 16.1528 & 51.0000 & 48.1870 & 10.4704 \\
$325$ & $[305, 350]$ & 42.7636 & 13.6664 & 0.7298 & 0.5299 & 43.4934 & 13.6767 & 44.0000 & {\pho\numRP{29.3945}{1}} & 6.9304 \\
$350$ & $[325, 370]$ & 26.0008 & 10.2234 & 0.2301 & 0.1548 & 26.2309 & 10.2246 & 37.0000 & {\pho\numRP{20.2227}{1}} & 4.2885 \\
$375$ & $[345, 395]$ & {\phoo\numRP{18.6252}{1}} & 9.7908 & 0.0759 & 0.0759 & {\pho\numRP{18.7012}{1}} & 9.7913 & 22.0000 & {\pho\numRP{12.5651}{1}} & 2.8120 \\
$400$ & $[375, 420]$ & {\phoo\pho\numRP{9.4616}{1}} & 7.6575 & 0.0256 & 0.0256 & {\phoo\numRP{9.4872}{1}} & 7.6576 & 5.0000 & {\phoo\numRP{8.0500}{1}} & 1.8309 \\
\bottomrule
\end{tabular}
\caption{Summary of the observed number of events in the data and the estimated number of signal and background events with total uncertainties (i.e. all listed uncertainties are the combined statistical and systematic uncertainties) that fall within each of the optimised mass windows in region $D2$. The total number of estimated background events in each window is the sum of the estimated background from the data-driven method and the $\ttbar$ simulation. The columns, from left to right indicate: $N_{B}^{\rm data\mbox{-}driven~est.}$, the data-driven background estimate; $N_{B}^{\ttbar~\rm est.}$, the background contribution from \ttbar; $N_{B}^{\rm tot.~est.}$, the total estimated background; $N_{\rm data}^{\rm obs.}$, the number of observed events in the data; and $N_S$, the number of expected signal events.}
\label{tab:results:windows}
\end{center}
\end{table}


Model-independent upper limits at 95\% confidence level (CL) on the number of beyond-the-SM (BSM) events for each signal region are derived using the \CLs prescription~\cite{HistFitter} and neglecting any possible contribution in the control regions. Dividing these by the integrated luminosity of the data sample provides upper limits on the visible BSM cross-section, $\sigma_{\rm vis.}$, which is defined as the product of acceptance ($A$), reconstruction efficiency ($\epsilon$), branching ratio ($\rm BR$), and production cross-section ($\sigma_\text{prod.}$). This search specifically targets low-mass $\stop\rightarrow\bbar\sbar$ decays, assuming $100\%$ BR. The resulting limits on the number of BSM events and on the visible signal cross-section are shown in \tabref{results:upperlimits}. The significance of an excess can be quantified by the probability ($p_0$) that a background-only experiment has at least as many events as observed.
This $p$-value is also reported for each region in \tabref{results:upperlimits}, where $p_0 = 1-{\rm CL}_b$ and ${\rm CL}_b$ is the confidence level observed for the background-only hypothesis. The $p$-value is truncated at 0.5 for any signal region where the observed number of events is less than the expected number.
\begin{table}
\begin{center}\renewcommand\arraystretch{1.6}
\sisetup{round-mode=figures, round-precision=2,
  retain-explicit-plus=true, group-digits = true}
\begin{tabular}{
 c|
 S[table-format=1.1, table-number-alignment=center, round-mode=figures, round-precision=2]|
 S[table-format=3.0, table-number-alignment=center, round-mode=places, round-precision=0]|
 S[table-number-alignment=right, round-mode=figures, round-precision=2]@{$\,$}
 S[table-number-alignment=left, round-mode=figures, round-precision=2,]|
 S[table-format=1.2, table-number-alignment=center, round-mode=figures, round-precision=2]
} 
\toprule
\multicolumn{6}{c}{Model-independent upper limits at 95\% CL} \\
\midrule
{Window [GeV]} & {$\sigma_{\rm vis.}$~[fb]} & {Observed $N_{\rm BSM}$} & \multicolumn{2}{c|}{Expected $N_{\rm BSM}$} & {$p_0$}  \\
\midrule
$[95, 115]$  & 5.82 & 101.3 & {\phoo\numRF{126.9}{3}}       & \numpmRF{+50.0}{-35.5}{2} & 0.5  \\
$[115, 135]$ & 7.03 & 122.3 & {\phoo\numRF{127.9}{3}}       & \numpmRF{+50.4}{-35.8}{2} & 0.5  \\
$[135, 165]$ & 8.36 & 145.4 & {\phoo\numRF{160.2}{3}}       & \numpmRF{+39.8}{-44.9}{2} & 0.5  \\
$[165, 190]$ & 8.39 & 146.0 & {\phoo\numRF{108.6}{3}}       & \numpmRF{+42.9}{-30.6}{2} & 0.19  \\
$[185, 210]$ & 5.89 & 102.5 & {\phoo\numRF{99.6 }{2}}       & \numpmRF{+39.2}{-27.9}{2} & 0.47  \\
$[210, 235]$ & 5.13 & 89.3  & {\phoo\pho\numRF{79.3 }{2}}   & \numpmRF{+31.3}{-22.2}{2} & 0.36  \\
$[235, 265]$ & 4.20 & 73.1  & {\phoo\pho\numRF{59.9 }{2}}   & \numpmRF{+23.7}{-16.9}{2} & 0.28  \\
$[260, 295]$ & 2.15 & 37.5  & {\phoo\pho\numRF{42.6 }{2}}   & \numpmRF{+17.0}{-12.0}{2} & 0.5  \\
$[280, 315]$ & 1.41 & 24.6  & {\phoo\pho\numRF{34.8 }{2}}   & \numpmRF{+13.8}{-10}{2}  & 0.5  \\
$[305, 350]$ & 1.74 & 30.2  & {\phoo\pho\numRF{29.9 }{2}}   & \numpmRF{+11.9}{-8}{2}  & 0.49  \\
$[325, 370]$ & 1.82 & {\phdoo\numRP{31.8}{1}} & {\phoo\phdoo\numRP{23.5 }{1}} & \numpmRF{+9.4}{-6.6}{2}   & 0.18  \\
$[345, 395]$ & 1.37 & {\phdoo\numRP{23.8}{1}} & {\phoo\phdoo\numRP{21.4 }{1}} & \numpmRF{+8.4}{-6.0}{2}   & 0.38  \\
$[375, 420]$ & 0.57 & {\phdoo\numRP{10.0}{1}} & {\phoo\phdoo\numRP{10.8}{1}}  & \numpmRF{+3.2}{-2.1}{2}   &  0.5  \\
\bottomrule
\end{tabular}
\end{center}
\caption[Breakdown of upper limits.]{
Left to right: mass window range, 95\% CL upper limits on the visible cross-section ($\sigma_{\rm vis.} = \langle A \times \epsilon \times {\rm BR} \times {\rm \sigma_\text{prod.}}\rangle$) and on the number of signal events (Observed $N_{\rm BSM}$).  The fourth column (Expected $N_{\rm BSM}$) shows the 95\% CL upper limit on the number of signal events, given the expected number (and $\pm 1\sigma$ excursions on the expectation) of background events. The last column indicates the discovery $p$-value, $p_0 = 1-{\rm CL}_b$, where ${\rm CL}_b$ is the confidence level observed for the background-only hypothesis. The $p$-value is truncated at 0.5 for any mass window where the observed number of events is less than the expected number.
\label{tab:results:upperlimits}}
\end{table}

\begin{figure}[!ht]
    \centering
    \includegraphics[width=0.85\columnwidth]{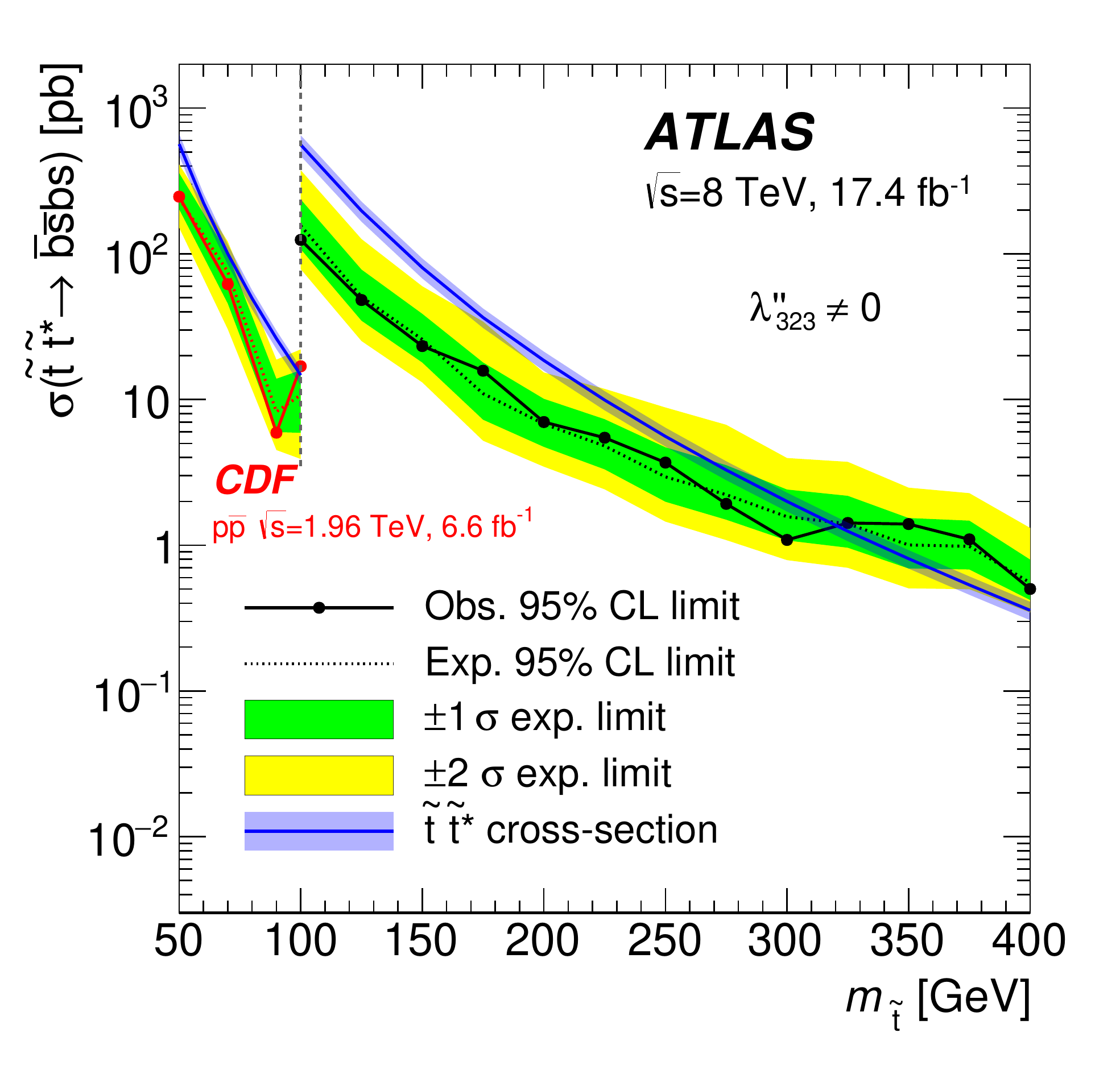}
    \caption{Observed and expected 95\% CL upper limits on the stop pair production cross-section as function of the stop mass. The solid line with big round markers shows the observed limit, the dotted line shows the expected exclusion limit, and the green and yellow bands represent the uncertainties on this limit. Limits from the CDF Collaboration are shown in red for $\mstop \leq 100 \gev$~\cite{Aaltonen:2013hya}. The blue line shows the theoretical signal cross-section and the blue band indicates the $\pm1 \sigma$ variations due to theoretical uncertainties on the signal production cross-section given by renormalisation and factorisation scale and PDF uncertainties. For this search the cross-section is calculated at NLO+NLL, whereas in the CDF paper the cross-section was calculated at NLO only.}
  \label{fig:results:limits}
\end{figure}

Exclusion limits are set on the signal model of interest. A profile likelihood ratio combining Poisson probabilities for signal and background is computed to determine the 95\% CL for compatibility of the data with the signal-plus-background hypothesis (\CLsb)~\cite{Cowan2011}. A similar calculation is performed for the background-only hypothesis (\CLb). From the ratio of these two quantities, the confidence level for the presence of signal (\CLs) is determined~\cite{HistFitter}. Systematic uncertainties are treated as nuisance parameters assuming Gaussian distributions and pseudo-experiments are used to evaluate the results. This procedure is implemented using a software framework for statistical data analysis,  HistFitter~\cite{Baak:2014wma}.
The observed and expected 95\% CL upper limits on the allowed cross-section are shown in \figref{results:limits}. For each simulated stop mass, the optimal mass window is chosen and the expected background yield is compared to the observed number of events in the mass window. Any potential signal contribution in the control regions from which the background estimates are derived is included as a systematic uncertainty on the background estimate. The size of the potential signal contribution in the control regions is shown for a few mass windows in \tabref{bkgdest:numbers}. Stops with masses between $100 \leq \mstop \leq 315\GeV$ are excluded at 95\% confidence level. 
All mass limits are quoted using the $\stop\stop^{*}$ signal production cross-section reduced by one standard deviation of the theory uncertainties.  

\FloatBarrier
\section{Conclusions}
\label{sec:conclusions}
This paper presents a search for direct pair production of light top squarks, decaying via an $R$-parity-violating coupling to $b$- and $s$-quarks. This leads to a final state characterised by two large-radius hadronic jets that each contain both decay products of the top squark. The search uses \intluminoerr of \sqseight proton--proton collision data collected with the ATLAS detector at the LHC. No deviation from the background prediction is observed, and  top squarks with masses between $100$ and $315\gev$ are excluded at 95\% confidence level.

\section{Acknowledgements}
\label{sec:ack}

We thank CERN for the very successful operation of the LHC, as well as the
support staff from our institutions without whom ATLAS could not be
operated efficiently.

We acknowledge the support of ANPCyT, Argentina; YerPhI, Armenia; ARC, Australia; BMWFW and FWF, Austria; ANAS, Azerbaijan; SSTC, Belarus; CNPq and FAPESP, Brazil; NSERC, NRC and CFI, Canada; CERN; CONICYT, Chile; CAS, MOST and NSFC, China; COLCIENCIAS, Colombia; MSMT CR, MPO CR and VSC CR, Czech Republic; DNRF and DNSRC, Denmark; IN2P3-CNRS, CEA-DSM/IRFU, France; GNSF, Georgia; BMBF, HGF, and MPG, Germany; GSRT, Greece; RGC, Hong Kong SAR, China; ISF, I-CORE and Benoziyo Center, Israel; INFN, Italy; MEXT and JSPS, Japan; CNRST, Morocco; FOM and NWO, Netherlands; RCN, Norway; MNiSW and NCN, Poland; FCT, Portugal; MNE/IFA, Romania; MES of Russia and NRC KI, Russian Federation; JINR; MESTD, Serbia; MSSR, Slovakia; ARRS and MIZ\v{S}, Slovenia; DST/NRF, South Africa; MINECO, Spain; SRC and Wallenberg Foundation, Sweden; SERI, SNSF and Cantons of Bern and Geneva, Switzerland; MOST, Taiwan; TAEK, Turkey; STFC, United Kingdom; DOE and NSF, United States of America. In addition, individual groups and members have received support from BCKDF, the Canada Council, CANARIE, CRC, Compute Canada, FQRNT, and the Ontario Innovation Trust, Canada; EPLANET, ERC, FP7, Horizon 2020 and Marie Sk{\l}odowska-Curie Actions, European Union; Investissements d'Avenir Labex and Idex, ANR, R{\'e}gion Auvergne and Fondation Partager le Savoir, France; DFG and AvH Foundation, Germany; Herakleitos, Thales and Aristeia programmes co-financed by EU-ESF and the Greek NSRF; BSF, GIF and Minerva, Israel; BRF, Norway; Generalitat de Catalunya, Generalitat Valenciana, Spain; the Royal Society and Leverhulme Trust, United Kingdom.

The crucial computing support from all WLCG partners is acknowledged
gratefully, in particular from CERN and the ATLAS Tier-1 facilities at
TRIUMF (Canada), NDGF (Denmark, Norway, Sweden), CC-IN2P3 (France),
KIT/GridKA (Germany), INFN-CNAF (Italy), NL-T1 (Netherlands), PIC (Spain),
ASGC (Taiwan), RAL (UK) and BNL (USA) and in the Tier-2 facilities
worldwide.

\clearpage

\bibliographystyle{bibtex/bst/atlasBibStyleWithTitle}
\bibliography{bibtex/bib/RPVStop,bibtex/bib/ATLAS,bibtex/bib/ConfNotes,bibtex/bib/PubNotes}

\newpage
\begin{flushleft}
{\Large The ATLAS Collaboration}

\bigskip

G.~Aad$^\textrm{\scriptsize 85}$,
B.~Abbott$^\textrm{\scriptsize 112}$,
J.~Abdallah$^\textrm{\scriptsize 150}$,
O.~Abdinov$^\textrm{\scriptsize 11}$,
B.~Abeloos$^\textrm{\scriptsize 116}$,
R.~Aben$^\textrm{\scriptsize 106}$,
M.~Abolins$^\textrm{\scriptsize 90}$,
O.S.~AbouZeid$^\textrm{\scriptsize 157}$,
H.~Abramowicz$^\textrm{\scriptsize 152}$,
H.~Abreu$^\textrm{\scriptsize 151}$
% ...
\bigskip
\\
% ...
$^{11}$ Institute of Physics, Azerbaijan Academy of Sciences, Baku, Azerbaijan\\
% ...
$^{85}$ CPPM, Aix-Marseille Universit{\'e} and CNRS/IN2P3, Marseille, France\\
% ...
$^{90}$ Department of Physics and Astronomy, Michigan State University, East Lansing MI, United States of America\\
% ...
$^{106}$ Nikhef National Institute for Subatomic Physics and University of Amsterdam, Amsterdam, Netherlands\\
% ...
$^{112}$ Homer L. Dodge Department of Physics and Astronomy, University of Oklahoma, Norman OK, United States of America\\
% ...
$^{116}$ LAL, Univ. Paris-Sud, CNRS/IN2P3, Universit{\'e} Paris-Saclay, Orsay, France\\
% ...
$^{150}$ Institute of Physics, Academia Sinica, Taipei, Taiwan\\
$^{151}$ Department of Physics, Technion: Israel Institute of Technology, Haifa, Israel\\
$^{152}$ Raymond and Beverly Sackler School of Physics and Astronomy, Tel Aviv University, Tel Aviv, Israel\\
% ...
$^{157}$ Department of Physics, University of Toronto, Toronto ON, Canada\\
% ...
\end{flushleft}


\end{document}